\documentclass[showpacs,aps,prd]{revtex4}
\input epsf

\begin{document}

\title{Doubly heavy baryons in QCD sum rules}
\author{Jian-Rong Zhang and Ming-Qiu Huang}
\affiliation{Department of Physics, National University of Defense
Technology, Hunan 410073, China}
\date{\today}

\begin{abstract}
The mass spectra of doubly heavy baryons are systematically calculated
in the framework of QCD sum rules. With a tentative
heavy-diquark--light-quark configuration, the interpolating currents
representing the doubly heavy baryons are proposed. Contributions of
the operators up to dimension six are included in operator product
expansion. The numerical results are compatible with other theoretical
predictions, which may support the $(QQ)-(q)$ structure of doubly heavy
baryons.
\end{abstract}
\pacs {14.20.-c, 11.55.Hx, 12.38.Lg}\maketitle
\section{Introduction}\label{sec1}
The investigation for the doubly heavy baryon containing double
heavy-quark $Q$ ($Q=c,b$) and a single light-quark $q$ ($q=u,d,s$)
may provide with valuable information for understanding the
nonperturbative QCD effects. SELEX Collaboration has reported the
first observation of a candidate for a double charm baryon,
$\Xi_{cc}^{+}$ \cite{ksi-cc1}, which was confirmed through the
measurement of a different weak decay mode later \cite{ksi-cc2}.
Additionally, BABAR Collaboration and Belle Collaboration have
reported that they have not observed any evidence of doubly charmed
baryons in $e^{+}e^{-}$ annihilations \cite{ksi-cc3}. The
feasibility of doubly heavy baryons studied at the Large Hadron
Collider (with the design luminosity values of
$\mathcal{L}=10^{34}~\mbox{cm}^{-2}\mbox{s}^{-1}$ and
$\sqrt{s}=14~\mbox{TeV}$) was presented \cite{production}. To
observe the doubly heavy baryons, it is necessary to give reliable
theoretical predictions of their properties. So far there have been
various approaches, by which the doubly heavy baryon masses been
calculated. Such as quark models \cite{quark model,quark
model1,quark model2,quark model3,quark model4,quark model5}, MIT bag
model \cite{MIT bag model}, and mass formulas \cite{mass formulas}.
However, the nonperturbative QCD which dominates the low energy
physics phenomena has not been fully understood in theory yet.

Another comprehensive and reliable way for evaluating the
nonperturbative effects is the QCD sum rule \cite{svzsum}. In this
method, instead of a model-dependent treatment in terms of
constituent quarks, hadrons are represented by their interpolating
quark currents taken at large virtualities. The correlation function
of these currents is introduced and treated in the framework of the
operator product expansion (OPE), where the short and long distance
quark-gluon interactions are separated. The former are calculated
using QCD perturbation theory, whereas the latter are parameterized
in terms of vacuum condensates. The result of the QCD calculation is
then matched, via dispersion relation, to calculate observable
characteristics of the hadronic ground state. Furthermore, the
interactions of quark-gluon currents with QCD vacuum fields
critically depend on the quantum numbers of these currents. However,
the accuracy of this method is limited, on the one hand, by the
approximations in the OPE of the correlation functions and, on the
other hand, by a very complicated and largely unknown structure of
the hadronic dispersion integrals. There have the non-relativistic
QCD sum rule \cite{NRQCDSR} and the full QCD one \cite{EBagan} been
explored to gain the masses of doubly heavy baryons. Nevertheless,
Ref. \cite{EBagan} has only studied some doubly heavy baryons.
Motivated by evaluating the mass spectra of them, the analysis of
the ground-state doubly heavy baryons via QCD sum rules would be
done in this work.

The content of this paper is as follows. In Sec \ref{sec2}, QCD sum
rules for the doubly heavy baryons are derived. Section \ref{sec3}
contains numerical analysis and some discussions.
\section{Doubly heavy baryon QCD sum rules}\label{sec2}
In a tentative picture for doubly heavy baryon $Q Q q$ system, the
low mass $q$ orbits the tightly bound $Q Q$ pair. The $(QQ)-(q)$
structure may be described similar to $\bar{Q}q$ mesons, where the
$Q Q$ pair plays the same role of the heavy antiquark $\bar{Q}$ in
$\bar{Q}q$. The study of such configuration can help one to adopt
the appropriate interpolating currents. For the ground states, the
currents are correlated with the spin-parity quantum numbers $0^{+}$
and $1^{+}$ for the heavy $Q Q$ diquark system, along with the light
quark $q$ forming the state with $J^{P}=\frac{1}{2}^{+}$ and the
pair of degenerate states. For the latter case, the $Q Q$ diquark
has spin $1$, and the spin of the third quark is either parallel,
$J^{P}=\frac{3}{2}^{+}$, or antiparallel, $J^{P}=\frac{1}{2}^{+}$,
to the diquark.  The choice of $\Gamma_{k}$ and $\Gamma_{k}^{'}$
matrices in baryonic currents may be determined according to the
rules in \cite{evs}. For the baryon with $J^{P}=\frac{3}{2}^{+}$,
the current may be gained using $SU(3)$ symmetry relations
\cite{Ioffe}. Consequently, the following forms of currents are
adopted in present work
\begin{eqnarray}
j_{\Xi_{QQ}}&=&\varepsilon_{abc}(Q_{a}^{T}C\Gamma_{k}Q_{b})\Gamma_{k}^{'}q_{c},\nonumber\\
j_{\Xi_{QQ}^{*}}&=&\varepsilon_{abc}\frac{1}{\sqrt{3}}[2(q_{a}^{T}C\Gamma_{k}Q_{b})\Gamma_{k}^{'}Q_{c}+(Q_{a}^{T}C\Gamma_{k}Q_{b})\Gamma_{k}^{'}q_{c}],\nonumber\\
j_{\Omega_{QQ}}&=&\varepsilon_{abc}(Q_{a}^{T}C\Gamma_{k}Q_{b})\Gamma_{k}^{'}s_{c},\nonumber\\
j_{\Omega_{QQ}^{*}}&=&\varepsilon_{abc}\frac{1}{\sqrt{3}}[2(s_{a}^{T}C\Gamma_{k}Q_{b})\Gamma_{k}^{'}Q_{c}+(Q_{a}^{T}C\Gamma_{k}Q_{b})\Gamma_{k}^{'}s_{c}],\nonumber\\
j_{\Xi_{QQ'}}&=&\varepsilon_{abc}(Q_{a}^{T}C\Gamma_{k}Q'_{b})\Gamma_{k}^{'}q_{c},\nonumber\\
j_{\Xi_{QQ'}^{*}}&=&\varepsilon_{abc}\frac{1}{\sqrt{3}}[(q_{a}^{T}C\Gamma_{k}Q_{b})\Gamma_{k}^{'}Q'_{c}+(q_{a}^{T}C\Gamma_{k}Q'_{b})\Gamma_{k}^{'}Q_{c}+(Q_{a}^{T}C\Gamma_{k}Q_{b})\Gamma_{k}^{'}q_{c}],\\
j_{\Omega_{QQ'}}&=&\varepsilon_{abc}(Q_{a}^{T}C\Gamma_{k}Q'_{b})\Gamma_{k}^{'}s_{c},\nonumber\\
j_{\Omega_{QQ'}^{*}}&=&\varepsilon_{abc}\frac{1}{\sqrt{3}}[(s_{a}^{T}C\Gamma_{k}Q_{b})\Gamma_{k}^{'}Q'_{c}+(s_{a}^{T}C\Gamma_{k}Q'_{b})\Gamma_{k}^{'}Q_{c}+(Q_{a}^{T}C\Gamma_{k}Q_{b})\Gamma_{k}^{'}s_{c}],\nonumber\\
j_{\Xi'_{QQ'}}&=&\varepsilon_{abc}(Q_{a}^{T}C\Gamma_{k}Q'_{b})\Gamma_{k}^{'}q_{c},\nonumber\\
j_{\Omega'_{QQ'}}&=&\varepsilon_{abc}(Q_{a}^{T}C\Gamma_{k}Q'_{b})\Gamma_{k}^{'}s_{c}.\nonumber
\end{eqnarray}
Here the index $T$ means matrix transposition, $C$ is the charge
conjugation matrix, $a$, $b$, and $c$ are color indices, $Q$ and
$Q'$ denote heavy quarks, and $q$ is $u$ or $d$. The choice of
$\Gamma_{k}$ and $\Gamma_{k}^{'}$ matrices are listed in TABLE
\ref{table:1}.
\begin{table}[htb!]\caption{The choice of $\Gamma_{k}$ and $\Gamma_{k}^{'}$ matrices in baryonic currents. The index $d$ in $S_{d}$, $L_{d}$, and $J_{d}^{P_{d}}$ means diquark. $\{QQ\}$ denotes the diquark in the axial
vector state and $[QQ]$ denotes the diquark in the scalar state.}
 \centerline{\begin{tabular}{ p{1.5cm} p{2.5cm} p{2.0cm} p{1cm} p{1cm} p{2.0cm} p{2.0cm} p{2.0cm}} \hline\hline
Baryon              & quark content        &$J^{P}$               &  $S_{d}$     &  $L_{d}$     &  $J_{d}^{P_{d}}$         &   $\Gamma_{k}$     &     $\Gamma_{k}^{'}$           \\
$\Xi_{QQ}$          &$\{QQ\}q$             &$\frac{1}{2}^{+} $    &      1       &      0       &        $1^{+}$           &   $\gamma_{\mu}$   &     $\gamma_{\mu}\gamma_{5}$   \\
\hline
$\Xi_{QQ}^{*}$      &$\{QQ\}q$             &$\frac{3}{2}^{+} $    &      1       &      0       &        $1^{+}$           &   $\gamma_{\mu}$   &     $1$                        \\
\hline
$\Omega_{QQ}$       &$\{QQ\}s$             &$\frac{1}{2}^{+} $    &      1       &      0       &        $1^{+}$           &   $\gamma_{\mu}$   &     $\gamma_{\mu}\gamma_{5}$   \\
\hline
$\Omega_{QQ}^{*}$   &$\{QQ\}s$             &$\frac{3}{2}^{+} $    &      1       &      0       &        $1^{+}$           &   $\gamma_{\mu}$   &     $1$                        \\
\hline
$\Xi_{QQ'}$         &$\{QQ'\}q$            &$\frac{1}{2}^{+} $    &      1       &      0       &        $1^{+}$           &   $\gamma_{\mu}$   &     $\gamma_{\mu}\gamma_{5}$   \\
\hline
$\Xi_{QQ'}^{*}$     &$\{QQ'\}q$            &$\frac{3}{2}^{+} $    &      1       &      0       &        $1^{+}$           &   $\gamma_{\mu}$   &     $1$                        \\
\hline
$\Omega_{QQ'}$      &$\{QQ'\}s$            &$\frac{1}{2}^{+} $    &      1       &      0       &        $1^{+}$           &   $\gamma_{\mu}$   &     $\gamma_{\mu}\gamma_{5}$   \\
\hline
$\Omega_{QQ'}^{*}$  &$\{QQ'\}s$            &$\frac{3}{2}^{+} $    &      1       &      0       &        $1^{+}$           &   $\gamma_{\mu}$   &     $1$                        \\
\hline
$\Xi_{QQ'}^{'}$     &$[QQ']q$              &$\frac{1}{2}^{+} $    &      0       &      0       &        $0^{+}$           &   $\gamma_{5}$     &     $1$                        \\
\hline
$\Omega_{QQ'}^{'}$  &$[QQ']s$              &$\frac{1}{2}^{+} $    &      0       &      0       &        $0^{+}$           &   $\gamma_{5}$     &     $1$                        \\
\hline\hline
\end{tabular}}
\label{table:1}
\end{table}

The QCD sum rules for the doubly heavy baryons are constructed from
the two-point correlator
\begin{eqnarray}\label{correlator}
\Pi(q^{2})=i\int
d^{4}x\mbox{e}^{iq.x}\langle0|T[j(x)\overline{j}(0)]|0\rangle.
\end{eqnarray}
Lorentz covariance implies that the two-point correlation function
has the form
\begin{eqnarray}
\Pi(q^{2})=\rlap/q\Pi_{1}(q^{2})+\Pi_{2}(q^{2}).
\end{eqnarray}
For each invariant function $\Pi_{1}$ and $\Pi_{2}$, a sum rule can
be obtained, which is given below. Phenomenologically, the
correlator can be expressed as a dispersion integral over a physical
spectral function
\begin{eqnarray}
\Pi(q^{2})=\lambda^{2}_H\frac{\rlap/q+M_{H}}{M_{H}^{2}-q^{2}}+\frac{1}{\pi}\int_{s_{0}}
^{\infty}ds\frac{\mbox{Im}\Pi^{\mbox{phen}}(s)}{s-q^{2}}+\mbox{subtractions},
\end{eqnarray}
where $M_{H}$ denotes the mass of the doubly heavy baryon. In
obtaining the above expression, the Dirac and Rarita-Schwinger
spinor sum relations,
\begin{eqnarray}
\sum_{s}N(q,s)\bar{N}(q,s)=\rlap/q+M_{H},
\end{eqnarray}
for spin-$\frac{1}{2}$ baryon, and
\begin{eqnarray}
\sum_{s}N_{\mu}(q,s)\bar{N}_{\nu}(q,s)=(\rlap/q+M_{H})(g_{\mu\nu}-\frac{1}{3}\gamma_{\mu}\gamma_{\nu}+\frac{q_{\mu}\gamma_{\nu}-q_{\nu}\gamma_{\mu}}{3M_{H}}-\frac{2q_{\mu}q_{\nu}}{3M_{H}^{2}}),
\end{eqnarray}
for spin-$\frac{3}{2}$ baryon, have been used. In the OPE side, one
works at leading order in $\alpha_{s}$ and considers condensates up
to dimension six. The $s$ quark is dealt as a light one and the
diagrams are considered up to order $m_{s}$. To keep the heavy-quark
mass finite, one uses the momentum-space expression for the
heavy-quark propagator. We follow Refs. \cite{mnielsen} and
calculate the light-quark part of the correlation function in the
coordinate space, which is then Fourier-transformed to the momentum
space in $D$ dimension. The resulting light-quark part is combined
with the heavy-quark part before it is dimensionally regularized at
$D=4$. The correlation function can be written in terms of a
dispersion relation as
\begin{eqnarray}
\Pi_{i}(q^{2})=\int_{(m_{Q}+m_{Q'})^{2}}^{\infty}ds\frac{\rho_{i}(s)}{s-q^{2}},~~i=1,2,~
m_{Q}=m_{Q'}~\mbox{or}~m_{Q}\neq m_{Q'},
\end{eqnarray}
where the spectral density is given by the imaginary part of the
correlation function
\begin{eqnarray}
\rho_{i}(s)=\frac{1}{\pi}\mbox{Im}\Pi_{i}^{\mbox{OPE}}(s).
\end{eqnarray}
After equating the two expressions for $\Pi(q^{2})$, assuming
quark-hadron duality, and making a Borel transform, the sum rules
can be written as
\begin{eqnarray}\label{sumrule1}
\lambda_{H}^{2}e^{-M_{H}^{2}/M^{2}}&=&\int_{(m_{Q}+m_{Q'})^{2}}^{s_{0}}ds\rho_{1}(s)e^{-s/M^{2}},
\end{eqnarray}
\begin{eqnarray}\label{sumrule2}
\lambda_{H}^{2}M_{H}e^{-M_{H}^{2}/M^{2}}&=&\int_{(m_{Q}+m_{Q'})^{2}}^{s_{0}}ds\rho_{2}(s)e^{-s/M^{2}}.
\end{eqnarray}
To eliminate the baryon coupling constant $\lambda_H$ and extract
the $M_{H}$, we take the derivative of Eq. (\ref{sumrule1}) with
respect to $-\frac{1}{M^{2}}$, divide the result by Eq.
(\ref{sumrule1}) itself, and similarly deal with Eq.
(\ref{sumrule2}) to yield
\begin{eqnarray}\label{sum rule q}
M_{H}^{2}&=&\int_{(m_{Q}+m_{Q'})^{2}}^{s_{0}}ds\rho_{1}(s)s
e^{-s/M^{2}}/
\int_{(m_{Q}+m_{Q'})^{2}}^{s_{0}}ds\rho_{1}(s)e^{-s/M^{2}},
\end{eqnarray}
\begin{eqnarray}\label{sum rule m}
M_{H}^{2}&=&\int_{(m_{Q}+m_{Q'})^{2}}^{s_{0}}ds\rho_{2}(s)s
e^{-s/M^{2}}/
\int_{(m_{Q}+m_{Q'})^{2}}^{s_{0}}ds\rho_{2}(s)e^{-s/M^{2}}.
\end{eqnarray}
The spectral densities should be distinguished for two kinds of
doubly heavy baryons, namely, containing the same or differently
heavy quarks.  Firstly, with
\begin{eqnarray}
\rho_{1}(s)&=&-\frac{3}{2^{4}\pi^{4}}\int_{\alpha_{min}}^{\alpha_{max}}\frac{d\alpha}{\alpha}\int_{\beta_{min}}^{1-\alpha}\frac{d\beta}{\beta}[\alpha\beta
s-(\alpha+\beta)m_{Q}^{2}]^{2}\\&&{}
+\frac{3}{2^{2}\pi^{4}}m_{Q}^{2}\int_{\alpha_{min}}^{\alpha_{max}}\frac{d\alpha}{\alpha}\int_{\beta_{min}}^{1-\alpha}\frac{d\beta}{\beta}(1-\alpha-\beta)[\alpha\beta
s-(\alpha+\beta)m_{Q}^{2}],\nonumber
\end{eqnarray}
\begin{eqnarray}
\rho_{2}(s)&=&\frac{3\langle\bar{q}q\rangle}{\pi^{2}}\int_{\alpha_{min}}^{\alpha_{max}}d\alpha\alpha(1-\alpha)s\\&&{}
-\frac{3\langle g\bar{q}\sigma\cdot G
q\rangle}{2\pi^{2}}\int_{\alpha_{min}}^{\alpha_{max}}d\alpha\alpha(1-\alpha),\nonumber
\end{eqnarray}
for $\Xi_{QQ}$,
\begin{eqnarray}
\rho_{1}(s)&=&-\frac{3}{2^{5}\pi^{4}}\int_{\alpha_{min}}^{\alpha_{max}}\frac{d\alpha}{\alpha}\int_{\beta_{min}}^{1-\alpha}\frac{d\beta}{\beta}[\alpha\beta
s-(\alpha+\beta)m_{Q}^{2}]^{2}\nonumber\\&&{}
-\frac{1}{2^{3}\pi^{4}}m_{Q}^{2}\int_{\alpha_{min}}^{\alpha_{max}}\frac{d\alpha}{\alpha}\int_{\beta_{min}}^{1-\alpha}\frac{d\beta}{\beta}(1-\alpha-\beta)[\alpha\beta
s-(\alpha+\beta)m_{Q}^{2}]\\&&{}
-\frac{\langle\bar{q}q\rangle}{3\pi^{2}}m_{Q}\sqrt{1-4m_{Q}^{2}/s}\nonumber\\&&{}
 -\frac{\langle
g^{2}G^{2}\rangle}{3\cdot2^{6}\pi^{4}}\sqrt{1-4m_{Q}^{2}/s},\nonumber\\
\rho_{2}(s)&=&-\frac{1}{2^{3}\pi^{4}}m_{Q}\int_{\alpha_{min}}^{\alpha_{max}}\frac{d\alpha}{\alpha^{2}}\int_{\beta_{min}}^{1-\alpha}\frac{d\beta}{\beta}[\alpha\beta
s-(\alpha+\beta)m_{Q}^{2}]^{2}\nonumber\\&&{}
-\frac{\langle\bar{q}q\rangle}{3\cdot2^{2}\pi^{2}}\int_{\alpha_{min}}^{\alpha_{max}}d\alpha[3\alpha(1-\alpha)s+8m_{Q}^{2}]\nonumber\\&&{}
-\frac{\langle g^{2}
G^{2}\rangle}{3\cdot2^{5}\pi^{4}}m_{Q}\int_{\alpha_{min}}^{\alpha_{max}}\frac{d\alpha}{\alpha^{2}}\int_{\beta_{min}}^{1-\alpha}d\beta\beta\\&&{}
-\frac{\langle
g^{2}G^{2}\rangle}{3\cdot2^{5}\pi^{4}}m_{Q}\sqrt{1-4m_{Q}^{2}/s}\nonumber\\&&{}
+\frac{\langle g\bar{q}\sigma\cdot G
q\rangle}{2^{3}\pi^{2}}\int_{\alpha_{min}}^{\alpha_{max}}d\alpha\alpha(1-\alpha),\nonumber
\end{eqnarray}
for $\Xi_{QQ}^{*}$,
\begin{eqnarray}
\rho_{1}(s)&=&-\frac{3}{2^{4}\pi^{4}}\int_{\alpha_{min}}^{\alpha_{max}}\frac{d\alpha}{\alpha}\int_{\beta_{min}}^{1-\alpha}\frac{d\beta}{\beta}[\alpha\beta
s-(\alpha+\beta)m_{Q}^{2}]^{2}\nonumber\\&&{}
+\frac{3}{2^{2}\pi^{4}}m_{Q}^{2}\int_{\alpha_{min}}^{\alpha_{max}}\frac{d\alpha}{\alpha}\int_{\beta_{min}}^{1-\alpha}\frac{d\beta}{\beta}(1-\alpha-\beta)[\alpha\beta
s-(\alpha+\beta)m_{Q}^{2}]\\&&{}
-\frac{5\langle\bar{s}s\rangle}{2^{3}\pi^{2}}m_{s}\int_{\alpha_{min}}^{\alpha_{max}}d\alpha\alpha(1-\alpha),\nonumber
\end{eqnarray}
\begin{eqnarray}
\rho_{2}(s)&=&-\frac{3}{2^{3}\pi^{4}}m_{s}\int_{\alpha_{min}}^{\alpha_{max}}d\alpha\frac{1}{\alpha(1-\alpha)}[\alpha(1-\alpha)s-m_{Q}^{2}]^{2}\nonumber\\&&{}
+\frac{3}{2\pi^{4}}m_{s}m_{Q}^{2}\int_{\alpha_{min}}^{\alpha_{max}}\frac{d\alpha}{\alpha}\int_{\beta_{min}}^{1-\alpha}\frac{d\beta}{\beta}[\alpha\beta
s-(\alpha+\beta)m_{Q}^{2}]\nonumber\\&&{}
+\frac{3\langle\bar{s}s\rangle}{\pi^{2}}\int_{\alpha_{min}}^{\alpha_{max}}d\alpha\alpha(1-\alpha)
s\\&&{} +\frac{\langle
g^{2}G^{2}\rangle}{2^{5}\pi^{4}}m_{s}\sqrt{1-4m_{Q}^{2}/s}\nonumber\\&&{}
-\frac{3\langle g\bar{s}\sigma\cdot G
s\rangle}{2\pi^{2}}\int_{\alpha_{min}}^{\alpha_{max}}d\alpha\alpha(1-\alpha),\nonumber
\end{eqnarray}
for $\Omega_{QQ}$, and
\begin{eqnarray}
\rho_{1}(s)&=&-\frac{3}{2^{5}\pi^{4}}\int_{\alpha_{min}}^{\alpha_{max}}\frac{d\alpha}{\alpha}\int_{\beta_{min}}^{1-\alpha}\frac{d\beta}{\beta}[\alpha\beta
s-(\alpha+\beta)m_{Q}^{2}]^{2}\nonumber\\&&{}
-\frac{1}{2\pi^{4}}m_{s}m_{Q}\int_{\alpha_{min}}^{\alpha_{max}}d\alpha\int_{\beta_{min}}^{1-\alpha}\frac{d\beta}{\beta}[\alpha\beta
s-(\alpha+\beta)m_{Q}^{2}]\nonumber\\&&{}
-\frac{1}{2^{3}\pi^{4}}m_{Q}^{2}\int_{\alpha_{min}}^{\alpha_{max}}\frac{d\alpha}{\alpha}\int_{\beta_{min}}^{1-\alpha}\frac{d\beta}{\beta}(1-\alpha-\beta)[\alpha\beta
s-(\alpha+\beta)m_{Q}^{2}]\\&&{}
-\frac{\langle\bar{s}s\rangle}{3\pi^{2}}m_{Q}\sqrt{1-4m_{Q}^{2}/s}\nonumber\\&&{}
+\frac{29\langle\bar{s}s\rangle}{3\cdot2^{3}\pi^{2}}m_{s}\int_{\alpha_{min}}^{\alpha_{max}}d\alpha\alpha(1-\alpha)\nonumber\\&&{}
 -\frac{\langle
g^{2}G^{2}\rangle}{3\cdot2^{6}\pi^{4}}\sqrt{1-4m_{Q}^{2}/s},\nonumber\\
\rho_{2}(s)&=&-\frac{1}{2^{3}\pi^{4}}m_{Q}\int_{\alpha_{min}}^{\alpha_{max}}\frac{d\alpha}{\alpha^{2}}\int_{\beta_{min}}^{1-\alpha}\frac{d\beta}{\beta}[\alpha\beta
s-(\alpha+\beta)m_{Q}^{2}]^{2}\nonumber\\&&{}
+\frac{1}{2^{5}\pi^{4}}m_{s}\int_{\alpha_{min}}^{\alpha_{max}}\frac{d\alpha}{\alpha(1-\alpha)}[\alpha(1-\alpha)s-m_{Q}^{2}]^{2}\nonumber\\&&{}
-\frac{5}{2^{3}\pi^{4}}m_{s}m_{Q}^{2}\int_{\alpha_{min}}^{\alpha_{max}}\frac{d\alpha}{\alpha}\int_{\beta_{min}}^{1-\alpha}\frac{d\beta}{\beta}[\alpha\beta
s-(\alpha+\beta)m_{Q}^{2}]\nonumber\\&&{}
-\frac{\langle\bar{s}s\rangle}{3\cdot2^{2}\pi^{2}}\int_{\alpha_{min}}^{\alpha_{max}}d\alpha[3\alpha(1-\alpha)s+8m_{Q}^{2}-4(1-\alpha)m_{s}m_{Q}]\\&&{}
-\frac{\langle g^{2}
G^{2}\rangle}{3\cdot2^{5}\pi^{4}}m_{Q}\int_{\alpha_{min}}^{\alpha_{max}}\frac{d\alpha}{\alpha^{2}}\int_{\beta_{min}}^{1-\alpha}d\beta\beta\nonumber\\&&{}
-\frac{\langle
g^{2}G^{2}\rangle}{3\cdot2^{5}\pi^{4}}(m_{Q}+m_{s})\sqrt{1-4m_{Q}^{2}/s}\nonumber\\&&{}
+\frac{\langle g\bar{s}\sigma\cdot G
s\rangle}{2^{3}\pi^{2}}\int_{\alpha_{min}}^{\alpha_{max}}d\alpha\alpha(1-\alpha),\nonumber
\end{eqnarray}
for $\Omega_{QQ}^{*}$. The integration limits are given by
$\alpha_{min}=(1-\sqrt{1-4m_{Q}^{2}/s})/2$,
$\alpha_{max}=(1+\sqrt{1-4m_{Q}^{2}/s})/2$, and $\beta_{min}=\alpha
m_{Q}^{2}/(s\alpha-m_{Q}^{2})$. Secondly, with
\begin{eqnarray}
\rho_{1}(s)&=&-\frac{3}{2^{4}\pi^{4}}\int_{\alpha_{min}}^{\alpha_{max}}\frac{d\alpha}{\alpha}\int_{\beta_{min}}^{1-\alpha}\frac{d\beta}{\beta}[\alpha\beta
s-\alpha m_{Q}^{2}-\beta m_{Q'}^{2}]^{2}\\&&{}
+\frac{3}{2^{2}\pi^{4}}m_{Q}m_{Q'}\int_{\alpha_{min}}^{\alpha_{max}}\frac{d\alpha}{\alpha}\int_{\beta_{min}}^{1-\alpha}\frac{d\beta}{\beta}(1-\alpha-\beta)[\alpha\beta
s-\alpha m_{Q}^{2}-\beta m_{Q'}^{2}],\nonumber
\end{eqnarray}
\begin{eqnarray}
\rho_{2}(s)&=&\frac{\langle\bar{q}q\rangle}{\pi^{2}}\int_{\alpha_{min}}^{\alpha_{max}}d\alpha[3\alpha(1-\alpha)s-2\alpha
m_{Q}^{2}-2(1-\alpha)m_{Q'}^{2}+2m_{Q}m_{Q'}]\\&&{} -\frac{3\langle
g\bar{q}\sigma\cdot G
q\rangle}{2\pi^{2}}\int_{\alpha_{min}}^{\alpha_{max}}d\alpha\alpha(1-\alpha),\nonumber
\end{eqnarray}
for $\Xi_{QQ^{'}}$,
\begin{eqnarray}
\rho_{1}(s)&=&-\frac{1}{2^{5}\pi^{4}}\int_{\alpha_{min}}^{\alpha_{max}}\frac{d\alpha}{\alpha}\int_{\beta_{min}}^{1-\alpha}\frac{d\beta}{\beta}[\alpha\beta
s-\alpha m_{Q}^{2}-\beta m_{Q'}^{2}]^{2}\nonumber\\&&{}
-\frac{1}{2^{3}\pi^{4}}m_{Q}m_{Q'}\int_{\alpha_{min}}^{\alpha_{max}}\frac{d\alpha}{\alpha}\int_{\beta_{min}}^{1-\alpha}\frac{d\beta}{\beta}(1-\alpha-\beta)[\alpha\beta
s-\alpha m_{Q}^{2}-\beta m_{Q'}^{2}]\\&&{}
-\frac{\langle\bar{q}q\rangle}{3\cdot2\pi^{2}}\int_{\alpha_{min}}^{\alpha_{max}}d\alpha[\alpha
m_{Q'}+(1-\alpha)m_{Q}]\nonumber\\&&{}
 -\frac{\langle
g^{2}G^{2}\rangle}{3\cdot2^{7}\pi^{4}}\sqrt{(s-m_{Q}^{2}+m_{Q'}^{2})^{2}-4m_{Q'}^{2}s}/s,\nonumber\\
\rho_{2}(s)&=&-\frac{1}{2^{5}\pi^{4}}m_{Q'}\int_{\alpha_{min}}^{\alpha_{max}}\frac{d\alpha}{\alpha^{2}}\int_{\beta_{min}}^{1-\alpha}\frac{d\beta}{\beta}[\alpha\beta
s-\alpha m_{Q}^{2}-\beta m_{Q'}^{2}]^{2}\nonumber\\&&{}
-\frac{1}{2^{5}\pi^{4}}m_{Q}\int_{\alpha_{min}}^{\alpha_{max}}\frac{d\alpha}{\alpha}\int_{\beta_{min}}^{1-\alpha}\frac{d\beta}{\beta^{2}}[\alpha\beta
s-\alpha m_{Q}^{2}-\beta m_{Q'}^{2}]^{2}\nonumber\\&&{}
-\frac{\langle\bar{q}q\rangle}{3\cdot2^{2}\pi^{2}}\int_{\alpha_{min}}^{\alpha_{max}}d\alpha[6m_{Q}m_{Q'}+3\alpha(1-\alpha)s-2\alpha
m_{Q}^{2}-2(1-\alpha)m_{Q'}^{2}]\nonumber\\&&{}
 -\frac{\langle g^{2}
G^{2}\rangle}{3\cdot2^{7}\pi^{4}}m_{Q'}\int_{\alpha_{min}}^{\alpha_{max}}\frac{d\alpha}{\alpha^{2}}\int_{\beta_{min}}^{1-\alpha}d\beta\beta\\&&{}
-\frac{\langle g^{2}
G^{2}\rangle}{3\cdot2^{7}\pi^{4}}m_{Q}\int_{\alpha_{min}}^{\alpha_{max}}d\alpha\alpha\int_{\beta_{min}}^{1-\alpha}\frac{d\beta}{\beta^{2}}\nonumber\\&&{}
-\frac{\langle
g^{2}G^{2}\rangle}{3\cdot2^{7}\pi^{4}}(m_{Q}+m_{Q'})\sqrt{(s-m_{Q}^{2}+m_{Q'}^{2})^{2}-4m_{Q'}^{2}s}/s\nonumber\\&&{}
+\frac{\langle g\bar{q}\sigma\cdot G
q\rangle}{2^{3}\pi^{2}}\int_{\alpha_{min}}^{\alpha_{max}}d\alpha\alpha(1-\alpha),\nonumber
\end{eqnarray}
for $\Xi_{QQ^{'}}^{*}$,
\begin{eqnarray}
\rho_{1}(s)&=&-\frac{3}{2^{4}\pi^{4}}\int_{\alpha_{min}}^{\alpha_{max}}\frac{d\alpha}{\alpha}\int_{\beta_{min}}^{1-\alpha}\frac{d\beta}{\beta}[\alpha\beta
s-\alpha m_{Q}^{2}-\beta m_{Q'}^{2}]^{2}\nonumber\\&&{}
+\frac{3}{2^{2}\pi^{4}}m_{Q}m_{Q'}\int_{\alpha_{min}}^{\alpha_{max}}\frac{d\alpha}{\alpha}\int_{\beta_{min}}^{1-\alpha}\frac{d\beta}{\beta}(1-\alpha-\beta)[\alpha\beta
s-\alpha m_{Q}^{2}-\beta m_{Q'}^{2}]\\&&{}
-\frac{5\langle\bar{s}s\rangle}{2^{3}\pi^{2}}m_{s}\int_{\alpha_{min}}^{\alpha_{max}}d\alpha\alpha(1-\alpha),\nonumber
\end{eqnarray}
\begin{eqnarray}
\rho_{2}(s)&=&-\frac{3}{2^{3}\pi^{4}}m_{s}\int_{\alpha_{min}}^{\alpha_{max}}d\alpha\frac{1}{\alpha(1-\alpha)}[\alpha(1-\alpha)s-\alpha
m_{Q}^{2}-(1-\alpha)m_{Q'}^{2}]^{2}\nonumber\\&&{}
+\frac{3}{2\pi^{4}}m_{s}m_{Q}m_{Q'}\int_{\alpha_{min}}^{\alpha_{max}}\frac{d\alpha}{\alpha}\int_{\beta_{min}}^{1-\alpha}\frac{d\beta}{\beta}[\alpha\beta
s-\alpha m_{Q}^{2}-\beta m_{Q'}^{2}]\nonumber\\&&{}
+\frac{\langle\bar{s}s\rangle}{\pi^{2}}\int_{\alpha_{min}}^{\alpha_{max}}d\alpha[3\alpha(1-\alpha)s-2\alpha
m_{Q}^{2}-2(1-\alpha)m_{Q'}^{2}+2m_{Q}m_{Q'}]\\&&{}
 +\frac{\langle
g^{2}G^{2}\rangle}{2^{5}\pi^{4}}m_{s}\sqrt{(s-m_{Q}^{2}+m_{Q'}^{2})^{2}-4m_{Q'}^{2}s}/s\nonumber\\&&{}
-\frac{3\langle g\bar{s}\sigma\cdot G
s\rangle}{2\pi^{2}}\int_{\alpha_{min}}^{\alpha_{max}}d\alpha\alpha(1-\alpha),\nonumber
\end{eqnarray}
for $\Omega_{QQ^{'}}$,
\begin{eqnarray}
\rho_{1}(s)&=&-\frac{1}{2^{5}\pi^{4}}\int_{\alpha_{min}}^{\alpha_{max}}\frac{d\alpha}{\alpha}\int_{\beta_{min}}^{1-\alpha}\frac{d\beta}{\beta}[\alpha\beta
s-\alpha m_{Q}^{2}-\beta m_{Q'}^{2}]^{2}\nonumber\\&&{}-
\frac{1}{2^{3}\pi^{4}}m_{Q}m_{Q'}\int_{\alpha_{min}}^{\alpha_{max}}\frac{d\alpha}{\alpha}\int_{\beta_{min}}^{1-\alpha}\frac{d\beta}{\beta}(1-\alpha-\beta)[\alpha\beta
s-\alpha m_{Q}^{2}-\beta m_{Q'}^{2}]\nonumber\\&&{}
-\frac{1}{2^{3}\pi^{4}}m_{s}m_{Q'}\int_{\alpha_{min}}^{\alpha_{max}}d\alpha\int_{\beta_{min}}^{1-\alpha}\frac{d\beta}{\beta}[\alpha\beta
s-\alpha m_{Q}^{2}-\beta m_{Q'}^{2}]\\&&{}
-\frac{1}{2^{3}\pi^{4}}m_{s}m_{Q}\int_{\alpha_{min}}^{\alpha_{max}}\frac{d\alpha}{\alpha}\int_{\beta_{min}}^{1-\alpha}d\beta[\alpha\beta
s-\alpha m_{Q}^{2}-\beta m_{Q'}^{2}]\nonumber\\&&{}
-\frac{\langle\bar{s}s\rangle}{3\cdot2^{4}\pi^{2}}\int_{\alpha_{min}}^{\alpha_{max}}d\alpha[8\alpha
m_{Q'}+8(1-\alpha)m_{Q}-17\alpha(1-\alpha)m_{s}]\nonumber\\&&{}
-\frac{\langle
g^{2}G^{2}\rangle}{3\cdot2^{7}\pi^{4}}\sqrt{(s-m_{Q}^{2}+m_{Q'}^{2})^{2}-4m_{Q'}^{2}s}/s,\nonumber\\
\rho_{2}(s)&=&-\frac{1}{2^{5}\pi^{4}}m_{Q'}\int_{\alpha_{min}}^{\alpha_{max}}\frac{d\alpha}{\alpha^{2}}\int_{\beta_{min}}^{1-\alpha}\frac{d\beta}{\beta}[\alpha\beta
s-\alpha m_{Q}^{2}-\beta m_{Q'}^{2}]^{2}\nonumber\\&&{}
-\frac{1}{2^{5}\pi^{4}}m_{Q}\int_{\alpha_{min}}^{\alpha_{max}}\frac{d\alpha}{\alpha}\int_{\beta_{min}}^{1-\alpha}\frac{d\beta}{\beta^{2}}[\alpha\beta
s-\alpha m_{Q}^{2}-\beta m_{Q'}^{2}]^{2}\nonumber\\&&{}
+\frac{1}{2^{5}\pi^{4}}m_{s}\int_{\alpha_{min}}^{\alpha_{max}}\frac{d\alpha}{\alpha(1-\alpha)}[\alpha(1-\alpha)s-\alpha
m_{Q}^{2}-(1-\alpha)m_{Q'}^{2}]^{2}\nonumber\\&&{}
-\frac{3}{2^{3}\pi^{4}}m_{s}m_{Q}m_{Q'}\int_{\alpha_{min}}^{\alpha_{max}}\frac{d\alpha}{\alpha}\int_{\beta_{min}}^{1-\alpha}\frac{d\beta}{\beta}[\alpha\beta
s-\alpha m_{Q}^{2}-\beta m_{Q'}^{2}]\\&&{}
-\frac{\langle\bar{s}s\rangle}{3\cdot2^{2}\pi^{2}}\int_{\alpha_{min}}^{\alpha_{max}}d\alpha[6m_{Q}m_{Q'}-(1-\alpha)m_{s}m_{Q'}-\alpha
m_{s}m_{Q}+3\alpha(1-\alpha)s-2\alpha
m_{Q}^{2}-2(1-\alpha)m_{Q'}^{2}]\nonumber\\&&{}
 -\frac{\langle g^{2}
G^{2}\rangle}{3\cdot2^{7}\pi^{4}}m_{Q'}\int_{\alpha_{min}}^{\alpha_{max}}\frac{d\alpha}{\alpha^{2}}\int_{\beta_{min}}^{1-\alpha}d\beta\beta\nonumber\\&&{}
-\frac{\langle g^{2}
G^{2}\rangle}{3\cdot2^{7}\pi^{4}}m_{Q}\int_{\alpha_{min}}^{\alpha_{max}}d\alpha\alpha\int_{\beta_{min}}^{1-\alpha}\frac{d\beta}{\beta^{2}}\nonumber\\&&{}
-\frac{\langle
g^{2}G^{2}\rangle}{3\cdot2^{7}\pi^{4}}(m_{Q'}+m_{Q}+m_{s})\sqrt{(s-m_{Q}^{2}+m_{Q'}^{2})^{2}-4m_{Q'}^{2}s}/s\nonumber\\&&{}
+\frac{\langle g\bar{s}\sigma\cdot G
s\rangle}{2^{3}\pi^{2}}\int_{\alpha_{min}}^{\alpha_{max}}d\alpha\alpha(1-\alpha),\nonumber
\end{eqnarray}
for $\Omega_{QQ^{'}}^{*}$,
\begin{eqnarray}
\rho_{1}(s)&=&-\frac{3}{2^{6}\pi^{4}}\int_{\alpha_{min}}^{\alpha_{max}}\frac{d\alpha}{\alpha}\int_{\beta_{min}}^{1-\alpha}\frac{d\beta}{\beta}[\alpha\beta
s-\alpha m_{Q}^{2}-\beta m_{Q'}^{2}]^{2}\\&&{}
+\frac{3}{2^{5}\pi^{4}}m_{Q}m_{Q'}\int_{\alpha_{min}}^{\alpha_{max}}\frac{d\alpha}{\alpha}\int_{\beta_{min}}^{1-\alpha}\frac{d\beta}{\beta}(1-\alpha-\beta)[\alpha\beta
s-\alpha m_{Q}^{2}-\beta m_{Q'}^{2}],\nonumber
\end{eqnarray}
\begin{eqnarray}
\rho_{2}(s)&=&\frac{\langle\bar{q}q\rangle}{2^{3}\pi^{2}}\int_{\alpha_{min}}^{\alpha_{max}}d\alpha[m_{Q}m_{Q'}+3\alpha(1-\alpha)s-2\alpha
m_{Q}^{2}-2(1-\alpha)m_{Q'}^{2}]\\&&{} -\frac{3\langle
g\bar{q}\sigma\cdot G
q\rangle}{2^{4}\pi^{2}}\int_{\alpha_{min}}^{\alpha_{max}}d\alpha\alpha(1-\alpha),\nonumber
\end{eqnarray}
for $\Xi_{QQ^{'}}^{'}$, and
\begin{eqnarray}
\rho_{1}(s)&=&-\frac{3}{2^{6}\pi^{4}}\int_{\alpha_{min}}^{\alpha_{max}}\frac{d\alpha}{\alpha}\int_{\beta_{min}}^{1-\alpha}\frac{d\beta}{\beta}[\alpha\beta
s-\alpha m_{Q}^{2}-\beta m_{Q'}^{2}]^{2}\nonumber\\&&{}
+\frac{3}{2^{5}\pi^{4}}m_{Q}m_{Q'}\int_{\alpha_{min}}^{\alpha_{max}}\frac{d\alpha}{\alpha}\int_{\beta_{min}}^{1-\alpha}\frac{d\beta}{\beta}(1-\alpha-\beta)[\alpha\beta
s-\alpha m_{Q}^{2}-\beta m_{Q'}^{2}]\\&&{}
-\frac{5\langle\bar{s}s\rangle}{2^{5}\pi^{2}}m_{s}\int_{\alpha_{min}}^{\alpha_{max}}d\alpha\alpha(1-\alpha),\nonumber
\end{eqnarray}
\begin{eqnarray}
\rho_{2}(s)&=&-\frac{3}{2^{6}\pi^{4}}m_{s}\int_{\alpha_{min}}^{\alpha_{max}}d\alpha\frac{1}{\alpha(1-\alpha)}[\alpha(1-\alpha)s-\alpha
m_{Q}^{2}-(1-\alpha)m_{Q'}^{2}]^{2}\nonumber\\&&{}
+\frac{3}{2^{5}\pi^{4}}m_{s}m_{Q}m_{Q'}\int_{\alpha_{min}}^{\alpha_{max}}\frac{d\alpha}{\alpha}\int_{\beta_{min}}^{1-\alpha}\frac{d\beta}{\beta}[\alpha\beta
s-\alpha m_{Q}^{2}-\beta m_{Q'}^{2}]\nonumber\\&&{}
+\frac{\langle\bar{s}s\rangle}{2^{3}\pi^{2}}\int_{\alpha_{min}}^{\alpha_{max}}d\alpha[m_{Q}m_{Q'}+3\alpha(1-\alpha)s-2\alpha
m_{Q}^{2}-2(1-\alpha)m_{Q'}^{2}]\\&&{}
 -\frac{\langle
g^{2}G^{2}\rangle}{2^{8}\pi^{4}}m_{s}\sqrt{(s-m_{Q}^{2}+m_{Q'}^{2})^{2}-4m_{Q'}^{2}s}/s\nonumber\\&&{}
 -\frac{3\langle g\bar{s}\sigma\cdot G
s\rangle}{2^{4}\pi^{2}}\int_{\alpha_{min}}^{\alpha_{max}}d\alpha\alpha(1-\alpha),\nonumber
\end{eqnarray}
for $\Omega_{QQ^{'}}^{'}$. The integration limits are given by
$\alpha_{min}=[s-m_{Q}^{2}+m_{Q'}^{2}-\sqrt{(s-m_{Q}^{2}+m_{Q'}^{2})^{2}-4m_{Q'}^{2}s}]/(2s)$,
$\alpha_{max}=[s-m_{Q}^{2}+m_{Q'}^{2}+\sqrt{(s-m_{Q}^{2}+m_{Q'}^{2})^{2}-4m_{Q'}^{2}s}]/(2s)$,
and $\beta_{min}=\alpha m_{Q}^{2}/(s\alpha-m_{Q'}^{2})$.
\section{Numerical analysis and discussions}\label{sec3}
In this part, the sum rule (\ref{sum rule q}) will be numerically
analyzed for better convergence in contrast with the sum rule
(\ref{sum rule m}) for this work. The input values are taken as
$m_{c}=1.25~\mbox{GeV}, m_{b}=4.20~\mbox{GeV}$,
$m_{s}=0.13~\mbox{GeV},
\langle\bar{q}q\rangle=-(0.23)^{3}~\mbox{GeV}^{3},
\langle\bar{s}s\rangle=0.8~\langle\bar{q}q\rangle,\langle
g\bar{q}\sigma\cdot G q\rangle=m_{0}^{2}~\langle\bar{q}q\rangle,
m_{0}^{2}=0.8~\mbox{GeV}^{2}$, and $\langle
g^{2}G^{2}\rangle=0.88~\mbox{GeV}^{4}$. Complying with the standard
procedure of sum rule analysis, the threshold $s_{0}$ and Borel
parameter $M^{2}$ are varied to find the optimal stability window,
in which the perturbative contribution should be larger than the
condensate contributions while the pole contribution larger than
continuum contribution. Thus, the regions of thresholds are taken as
those presented in the figure captions, with
$M^{2}=3.5\sim5.0~\mbox{GeV}^{2}$ for $\Xi_{cc}$, $\Xi_{cc}^{*}$,
$\Omega_{cc}$, and $\Omega_{cc}^{*}$,
$M^{2}=7.5\sim9.0~\mbox{GeV}^{2}$ for $\Xi_{cb}$, $\Xi_{cb}^{*}$,
$\Omega_{cb}$, $\Omega_{cb}^{*}$, $\Xi_{cb}^{'}$, and
$\Omega_{cb}^{'}$, and $M^{2}=9.5\sim11.0~\mbox{GeV}^{2}$ for
$\Xi_{bb}$, $\Xi_{bb}^{*}$, $\Omega_{bb}$, and $\Omega_{bb}^{*}$,
respectively. The corresponding Borel curves are exhibited in Figs.
1-7. In Table \ref{table:2}, the numerical results are collected,
together with other theoretical predictions. It is worth noting that
uncertainty in our results are merely owing to the sum rule windows
(variation of the threshold $s_{0}$ and Borel parameter $M^{2}$),
not involving the ones from the variation of quark masses and QCD
parameters. Although the values of masses are in agrement with other
theoretical predictions, some of the absolute differences from them
are not small, for instance, the masses of $\Xi_{cc}$,
$\Omega_{cc}$, and $\Xi_{cb}^{*}$, whereas, in the tolerable ranges
of the sum rule method accuracy (the upper bound of relative
accuracy may approximate to $30\%$). Visually, the Borel curves for
$\Xi_{cc}$, $\Omega_{cc}$, and $\Xi_{cb}^{*}$ are not very flat, but
it is difficult to find much better sum rule windows. That's
probably because the condensate contributions for them, which may
play an important role in stabilizing the Borel curves, nearly
vanished or are small. The stability of these three curves might be
improved by including some higher dimension condensate
contributions.

\begin{figure}[htb!]
\centerline{\epsfysize=3.7truecm
\epsfbox{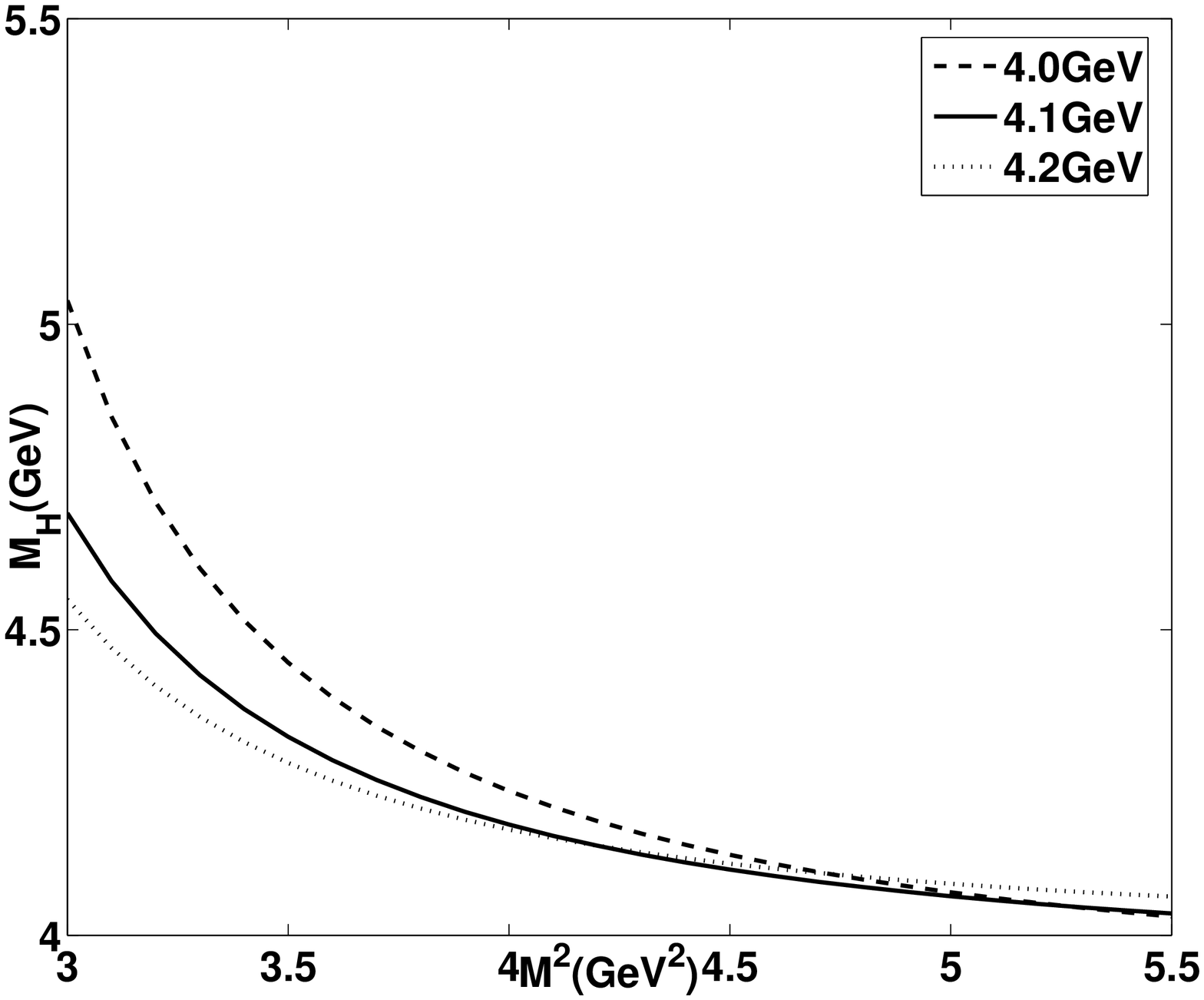}\epsfysize=3.7truecm
\epsfbox{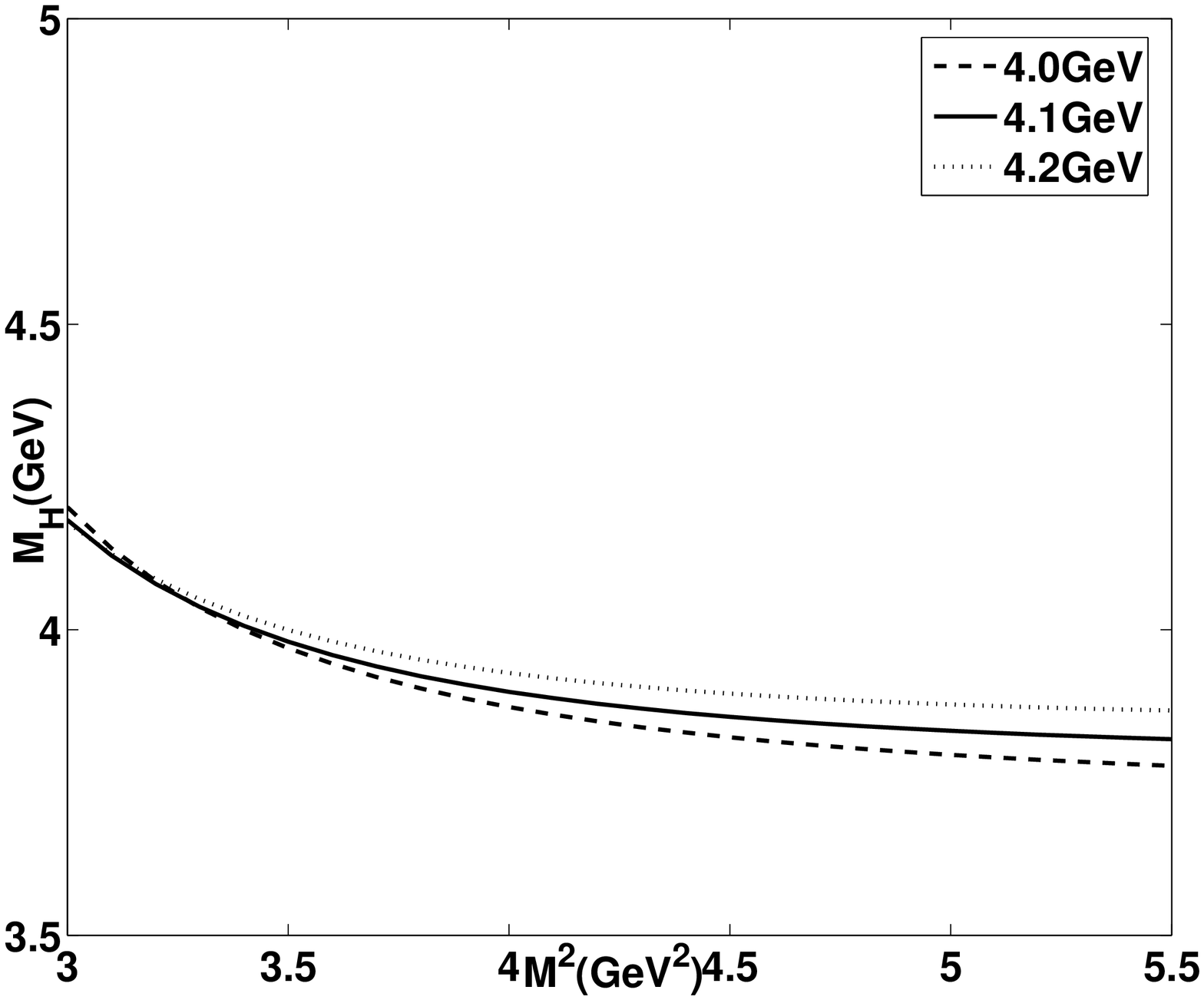}}\caption{The dependence on $M^2$ for the
masses of $\Xi_{cc}$ and $\Xi_{cc}^{*}$ from sum rule (\ref{sum rule
q}). The continuum thresholds are both taken as
$\sqrt{s_0}=4.0\sim4.2~\mbox{GeV}$.} \label{fig:1}
\end{figure}

\begin{figure}
\centerline{\epsfysize=3.7truecm
\epsfbox{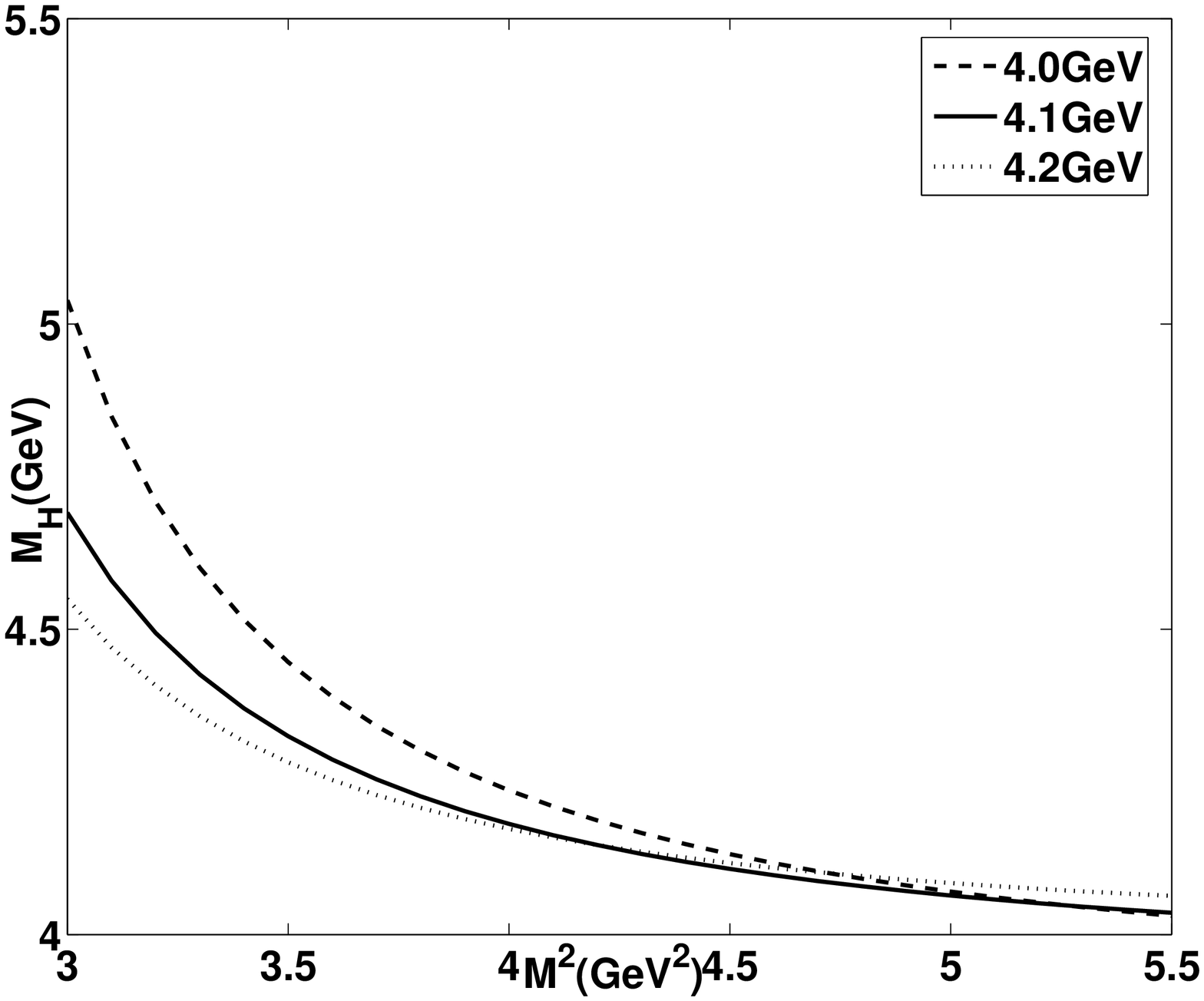}\epsfysize=3.7truecm
\epsfbox{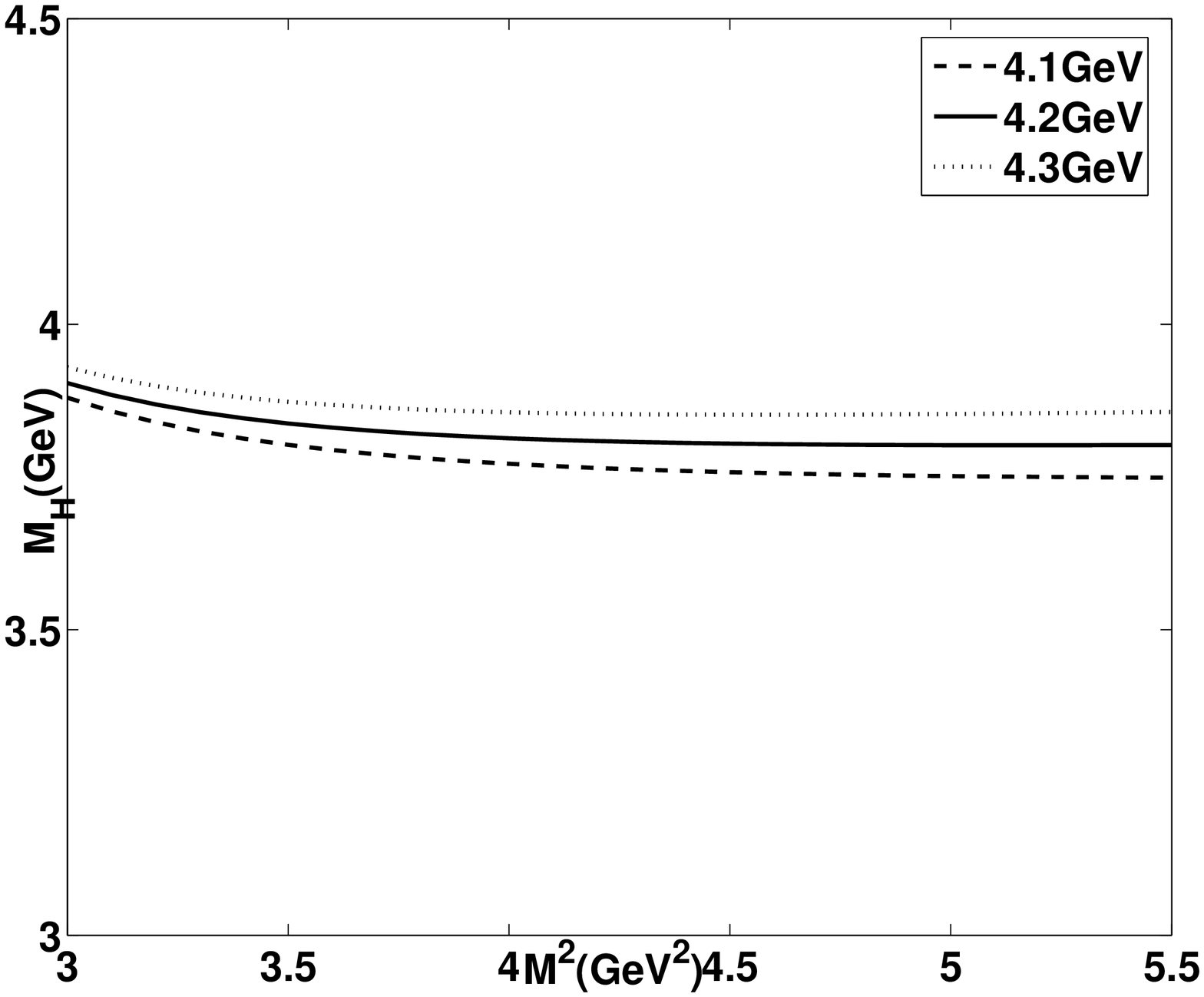}}\caption{The dependence on $M^2$ for the
masses of $\Omega_{cc}$ and $\Omega_{cc}^{*}$ from sum rule
(\ref{sum rule q}). The continuum thresholds are taken as
$\sqrt{s_0}=4.0\sim4.2~\mbox{GeV}$ and
$\sqrt{s_0}=4.1\sim4.3~\mbox{GeV}$.} \label{fig:2}
\end{figure}

\begin{figure}
\centerline{\epsfysize=3.7truecm
\epsfbox{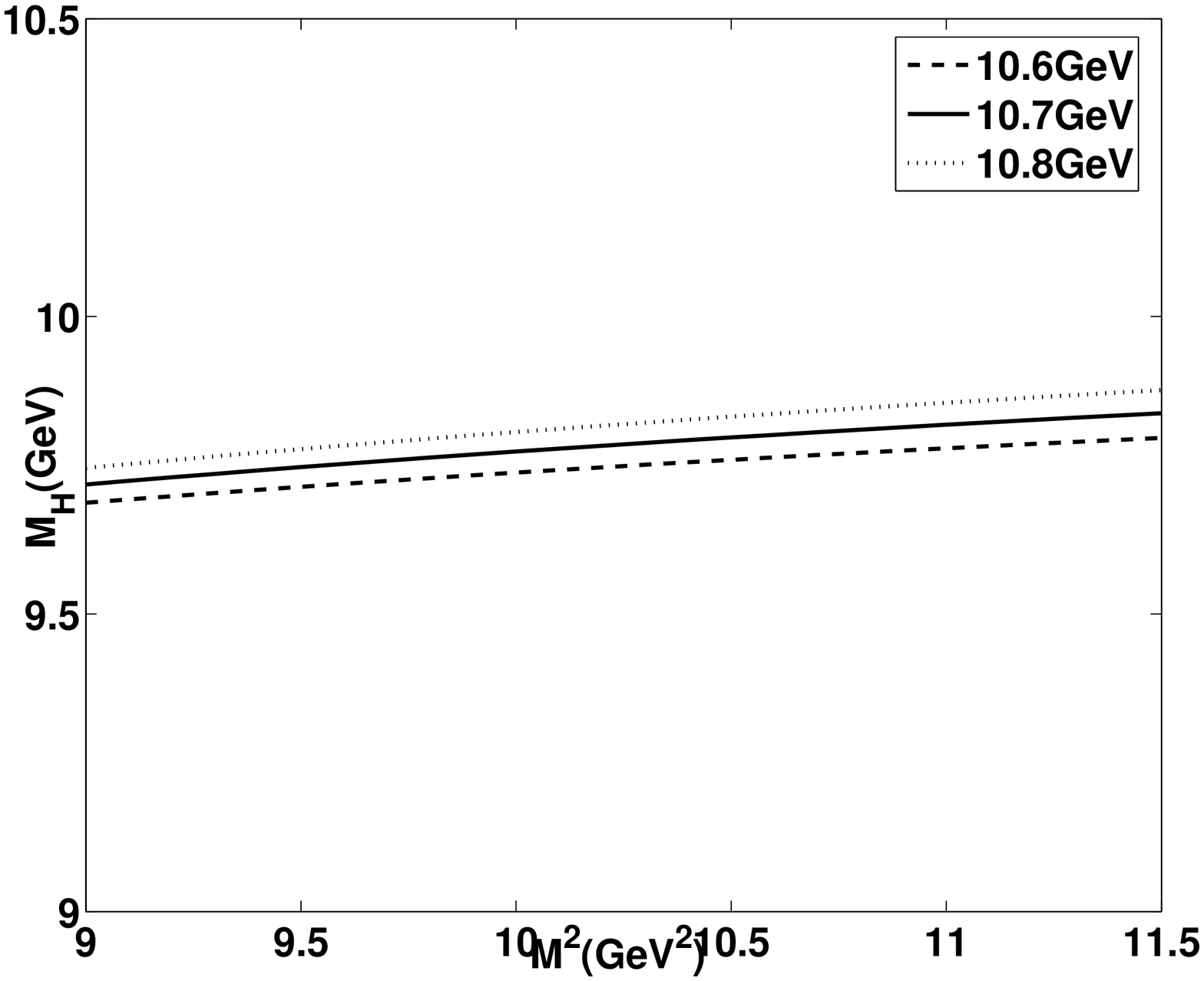}\epsfysize=3.7truecm
\epsfbox{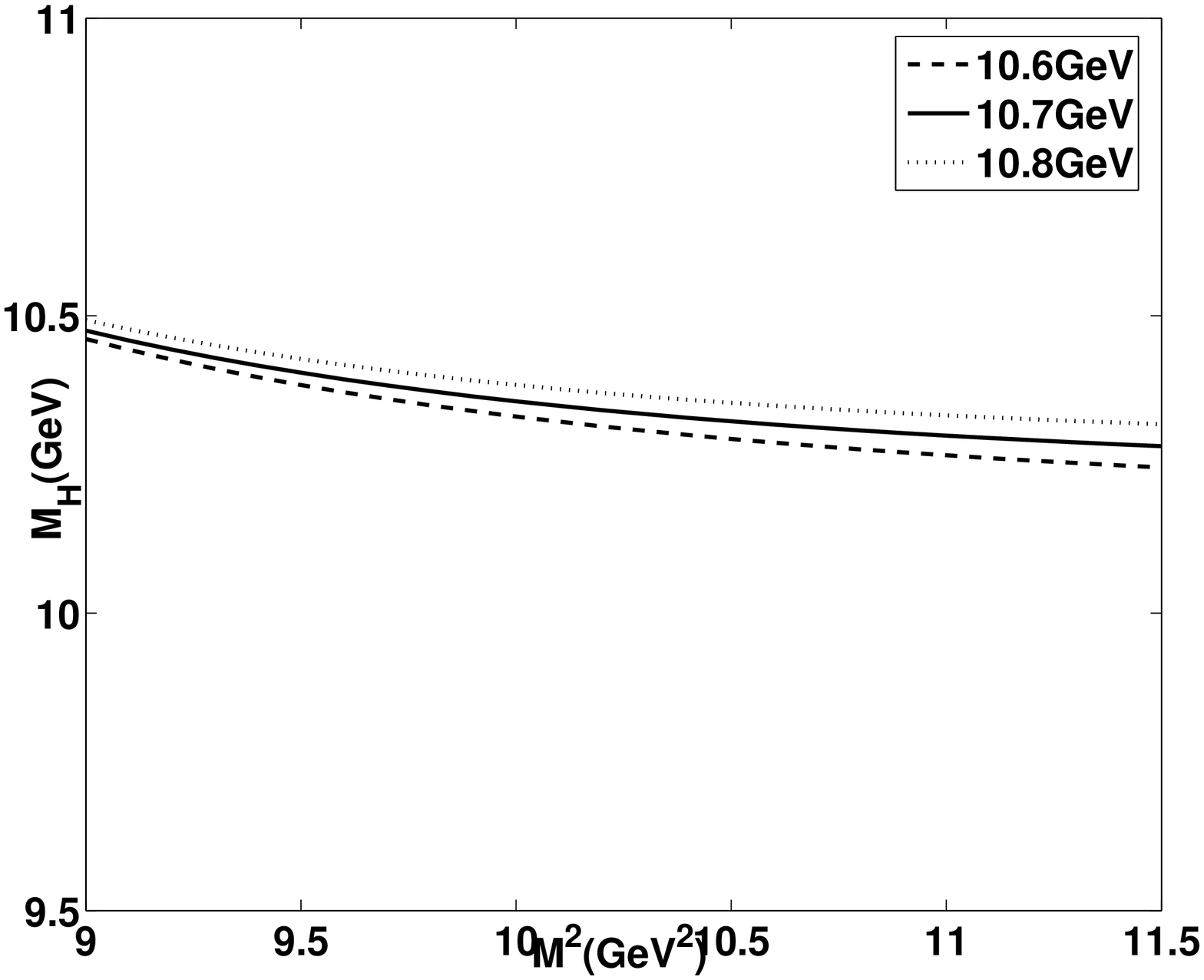}}\caption{The dependence on $M^2$ for the
masses of $\Xi_{bb}$ and $\Xi_{bb}^{*}$ from sum rule (\ref{sum rule
q}). The continuum thresholds are both taken as
$\sqrt{s_0}=10.6\sim10.8~\mbox{GeV}$.} \label{fig:3}
\end{figure}

\begin{figure}
\centerline{\epsfysize=3.7truecm
\epsfbox{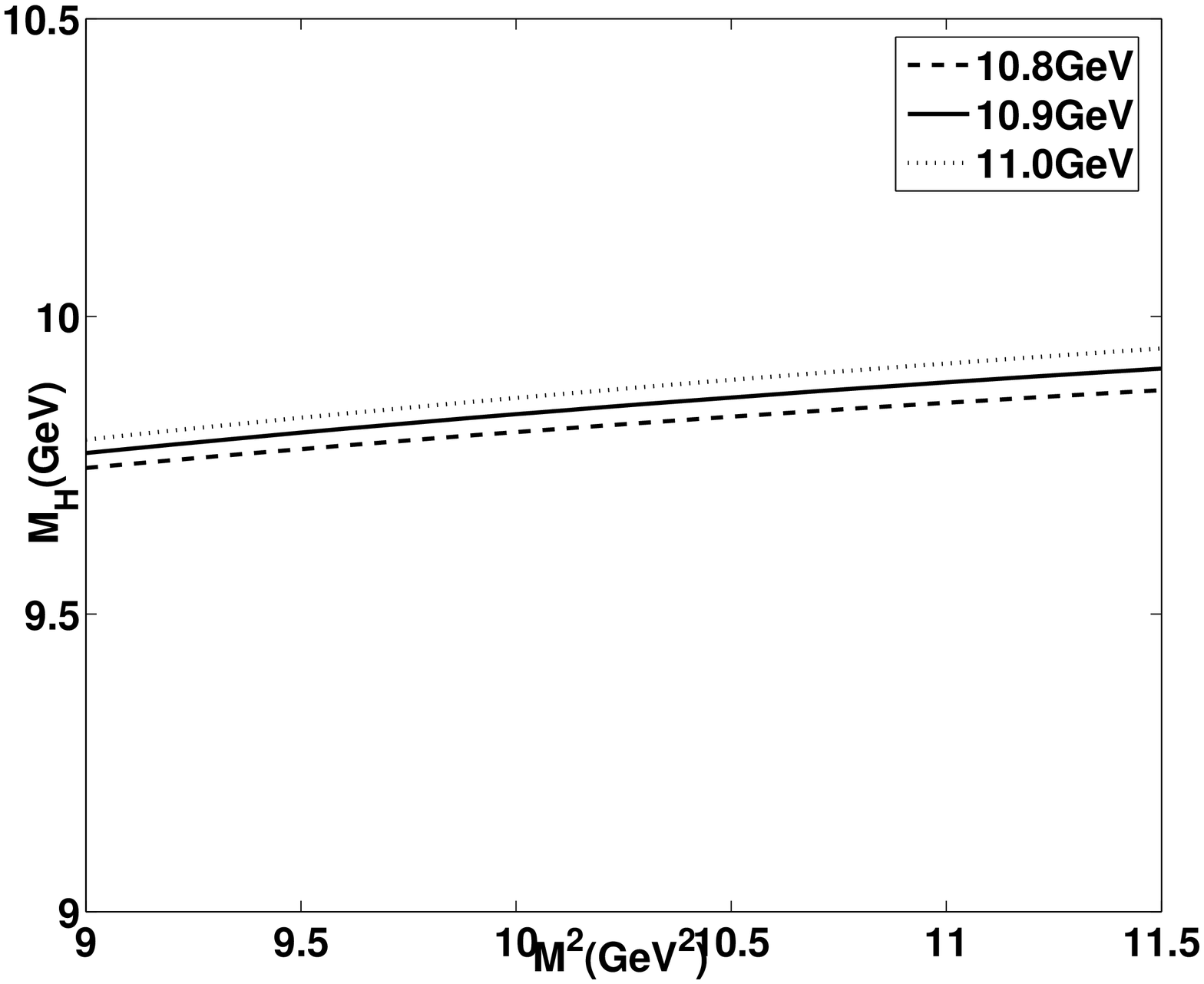}\epsfysize=3.7truecm
\epsfbox{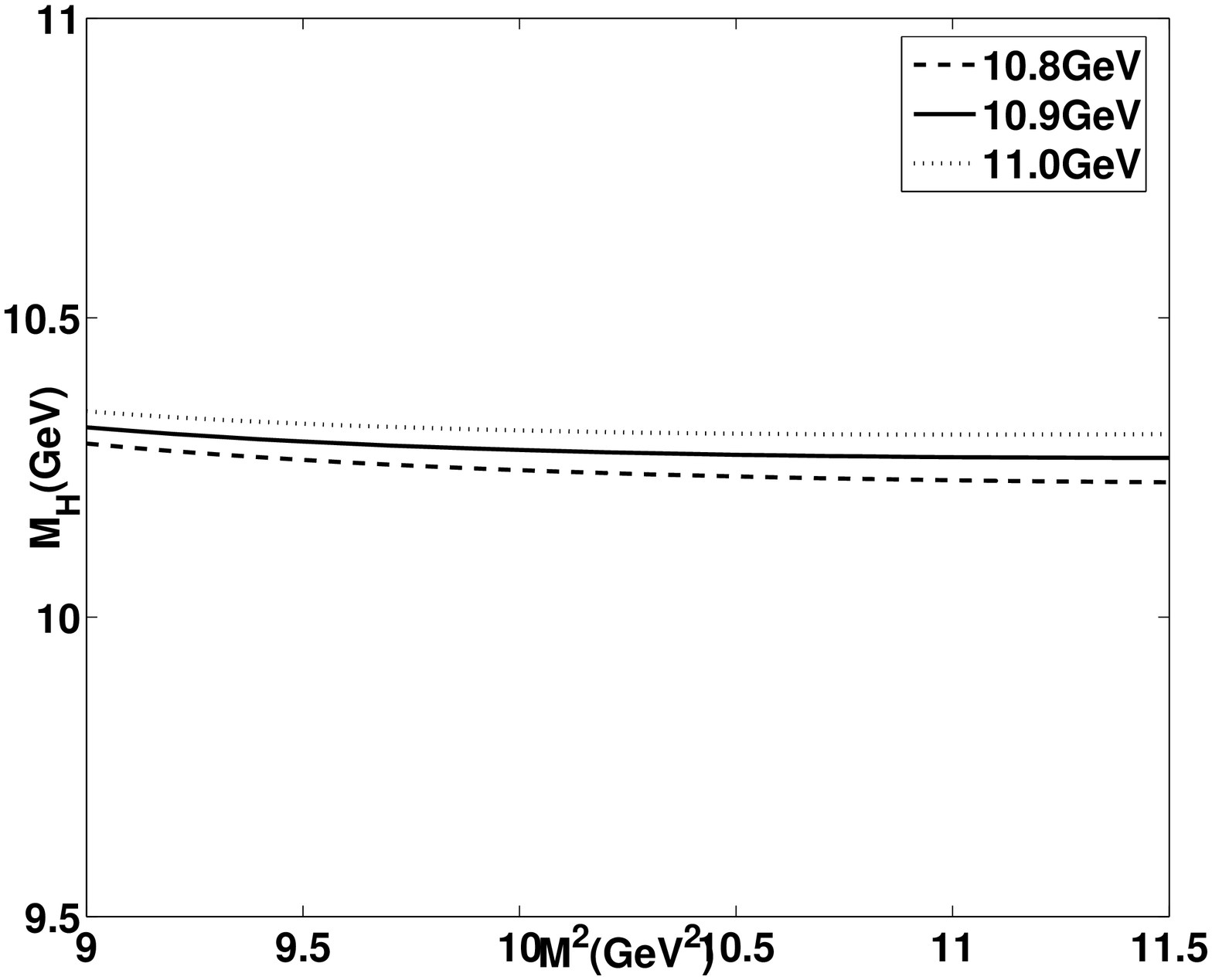}}\caption{The dependence on $M^2$ for the
masses of $\Omega_{bb}$ and $\Omega_{bb}^{*}$ from sum rule
(\ref{sum rule q}). The continuum thresholds are both taken as
$\sqrt{s_0}=10.8\sim11.0~\mbox{GeV}$.} \label{fig:4}
\end{figure}

\begin{figure}
\centerline{\epsfysize=3.7truecm
\epsfbox{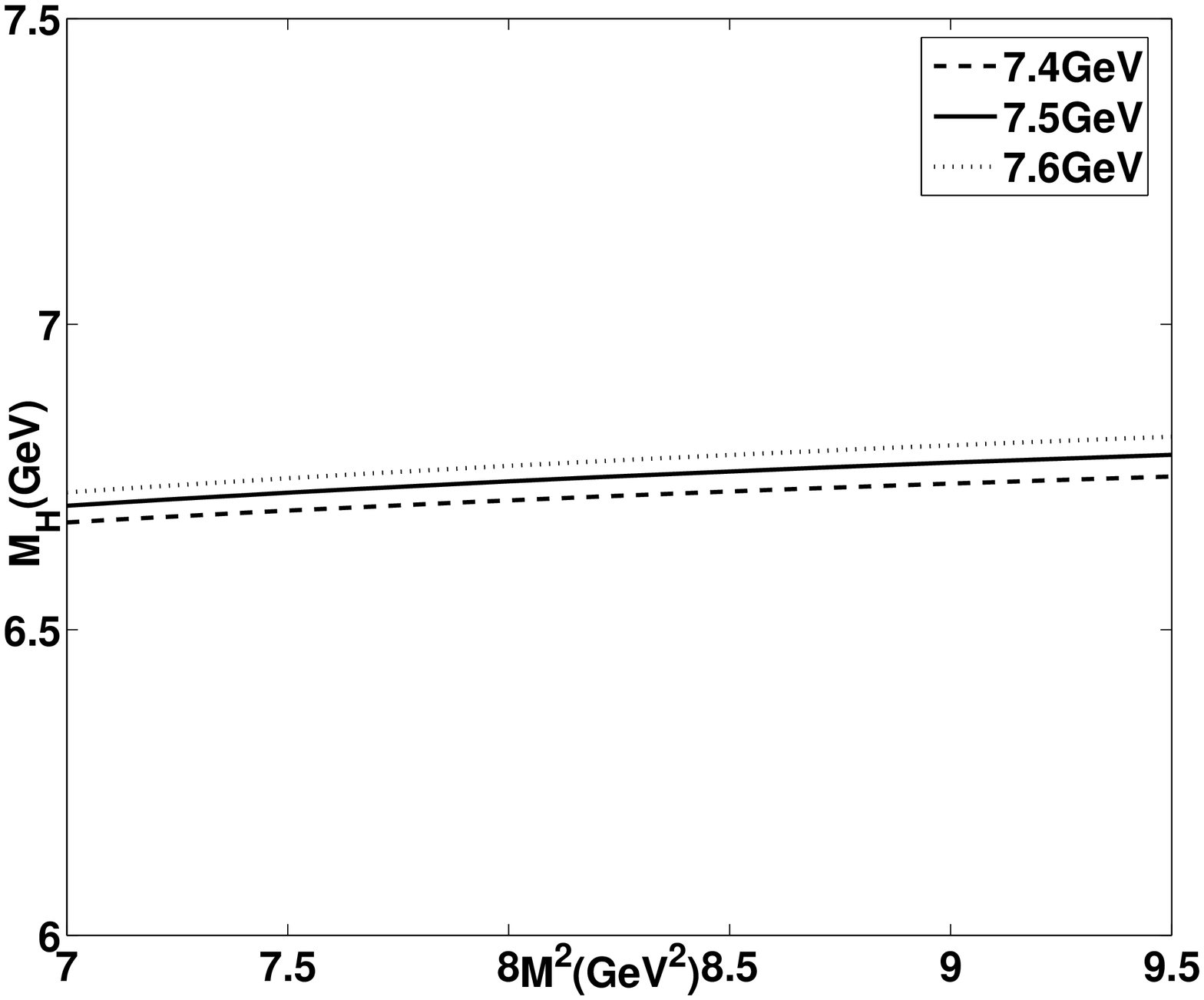}\epsfysize=3.7truecm
\epsfbox{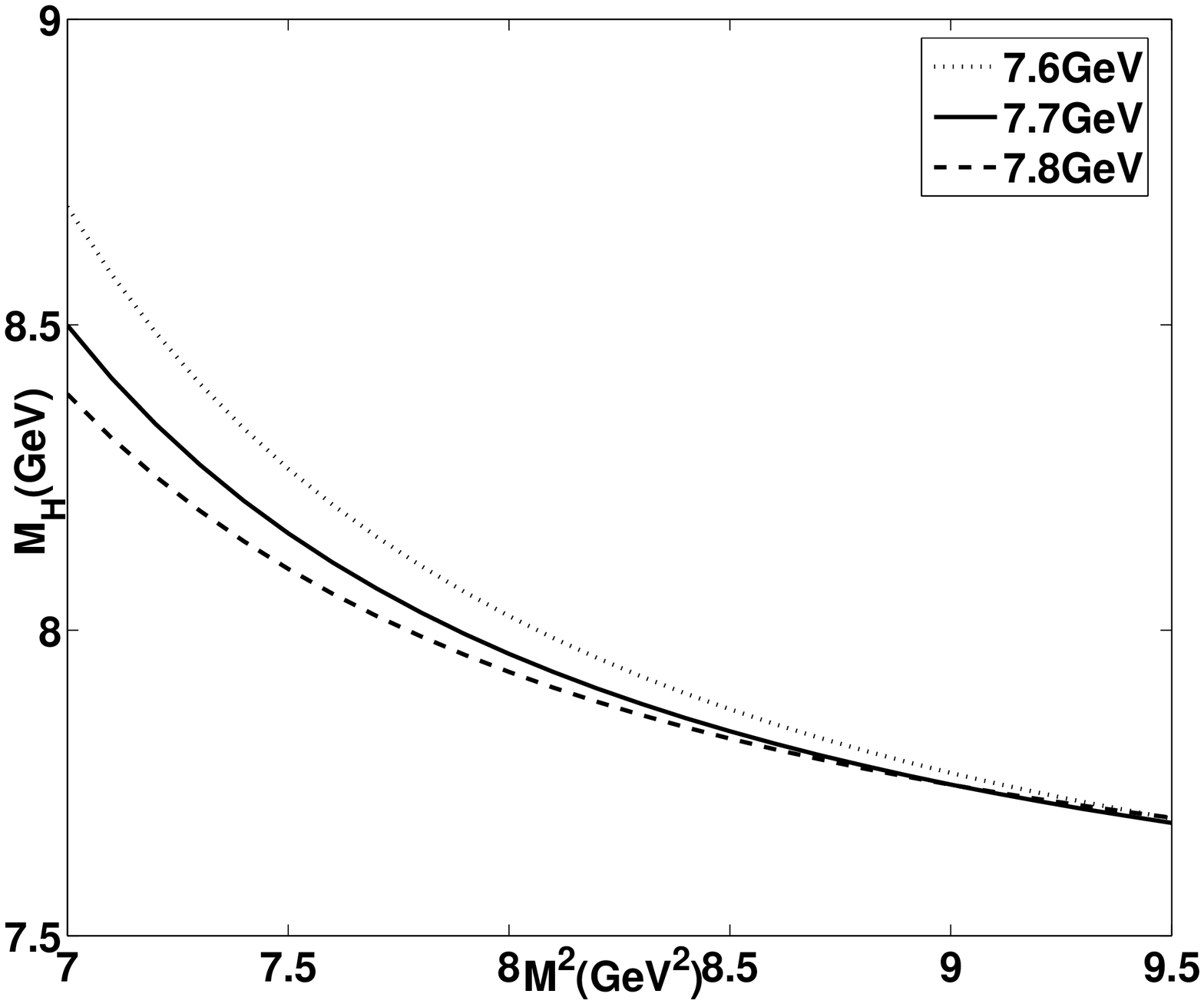}}\caption{The dependence on $M^2$ for the
masses of $\Xi_{cb}$ and $\Xi_{cb}^{*}$ from sum rule (\ref{sum rule
q}). The continuum thresholds are taken as
$\sqrt{s_0}=7.4\sim7.6~\mbox{GeV}$ and
$\sqrt{s_0}=7.6\sim7.8~\mbox{GeV}$.} \label{fig:5}
\end{figure}

\begin{figure}
\centerline{\epsfysize=3.7truecm
\epsfbox{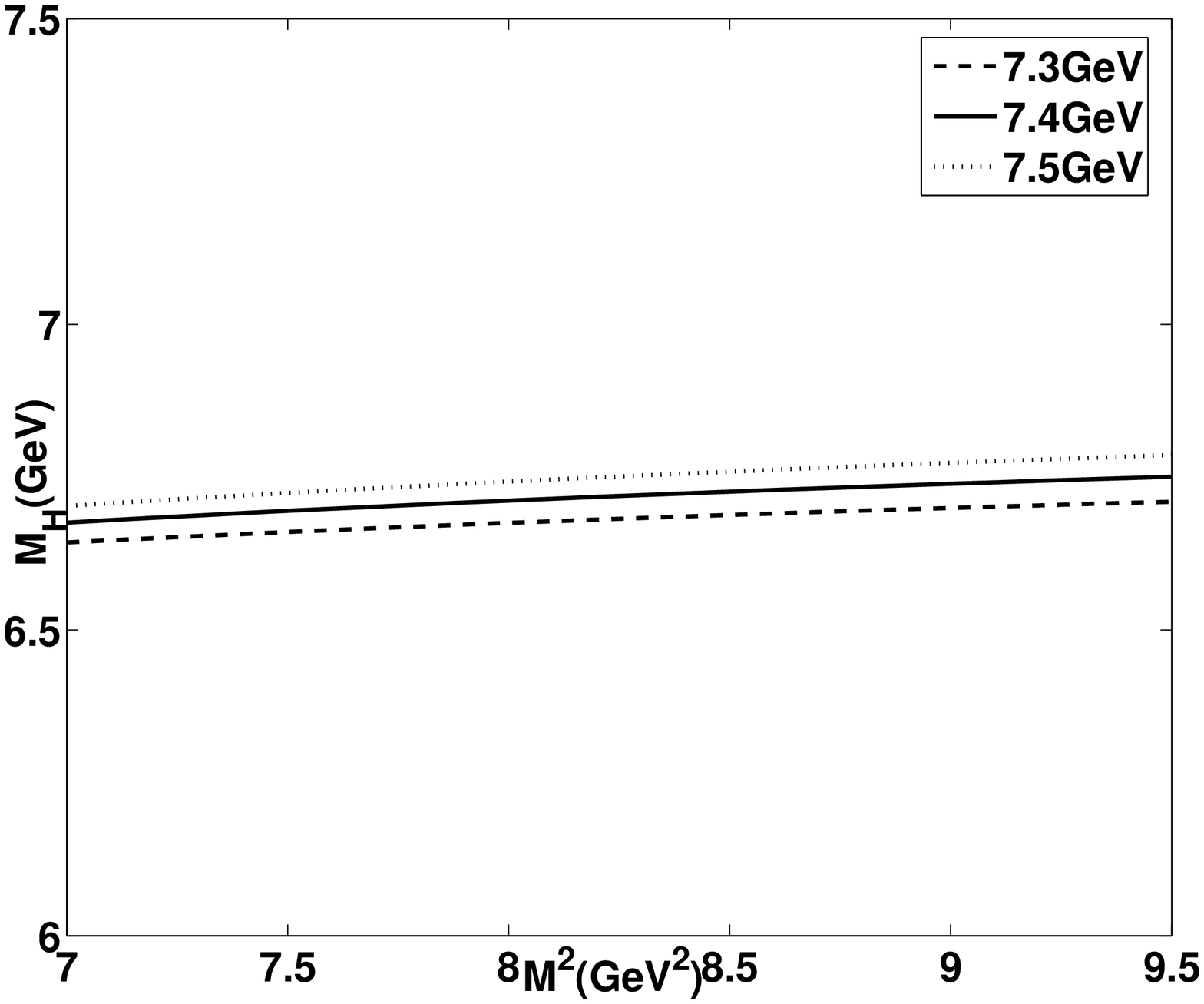}\epsfysize=3.7truecm
\epsfbox{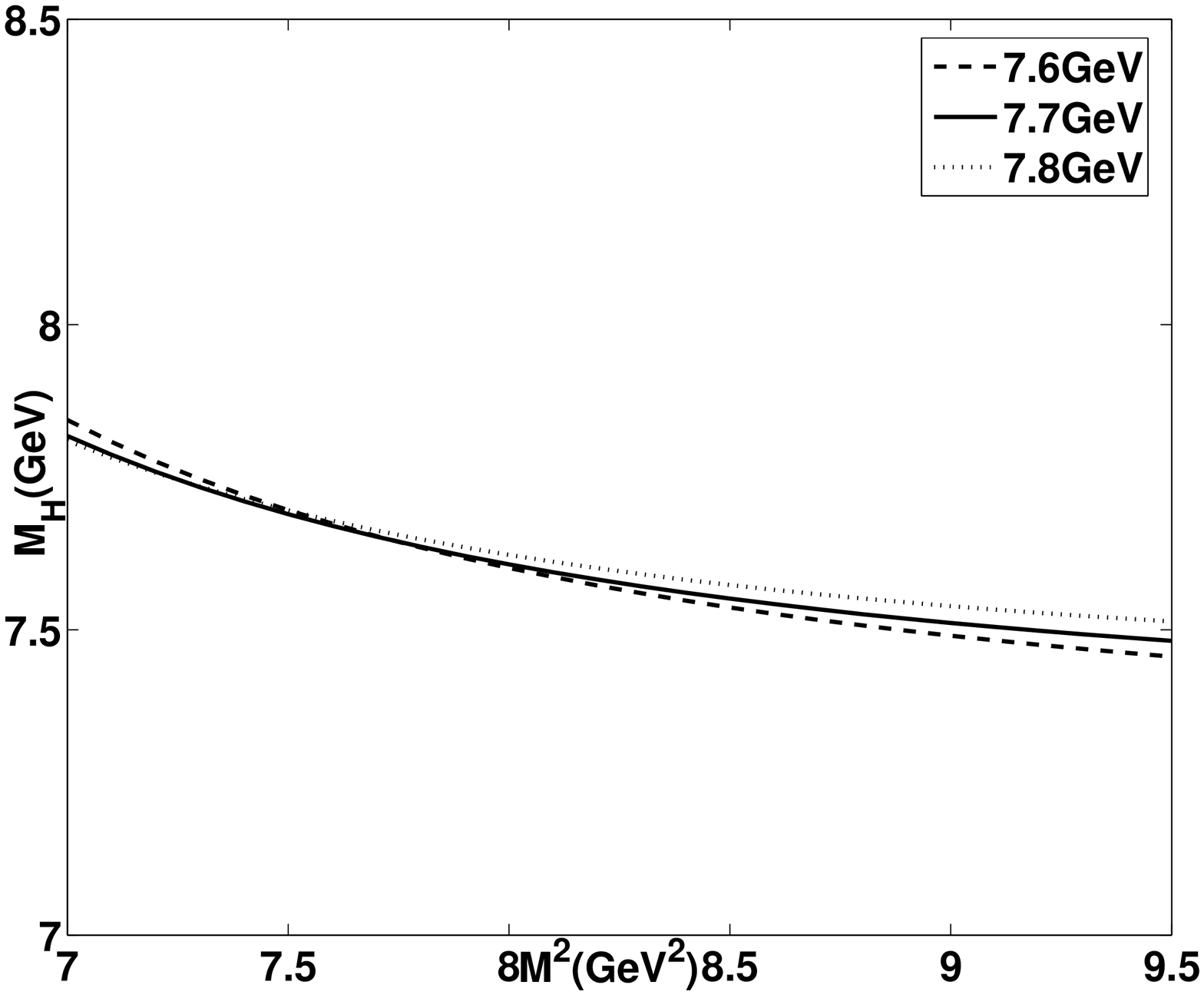}}\caption{The dependence on $M^2$ for the
masses of $\Omega_{cb}$ and $\Omega_{cb}^{*}$ from sum rule
(\ref{sum rule q}). The continuum thresholds are taken as
$\sqrt{s_0}=7.3\sim7.5~\mbox{GeV}$ and
$\sqrt{s_0}=7.6\sim7.8~\mbox{GeV}$.} \label{fig:6}
\end{figure}

\begin{figure}
\centerline{\epsfysize=3.7truecm
\epsfbox{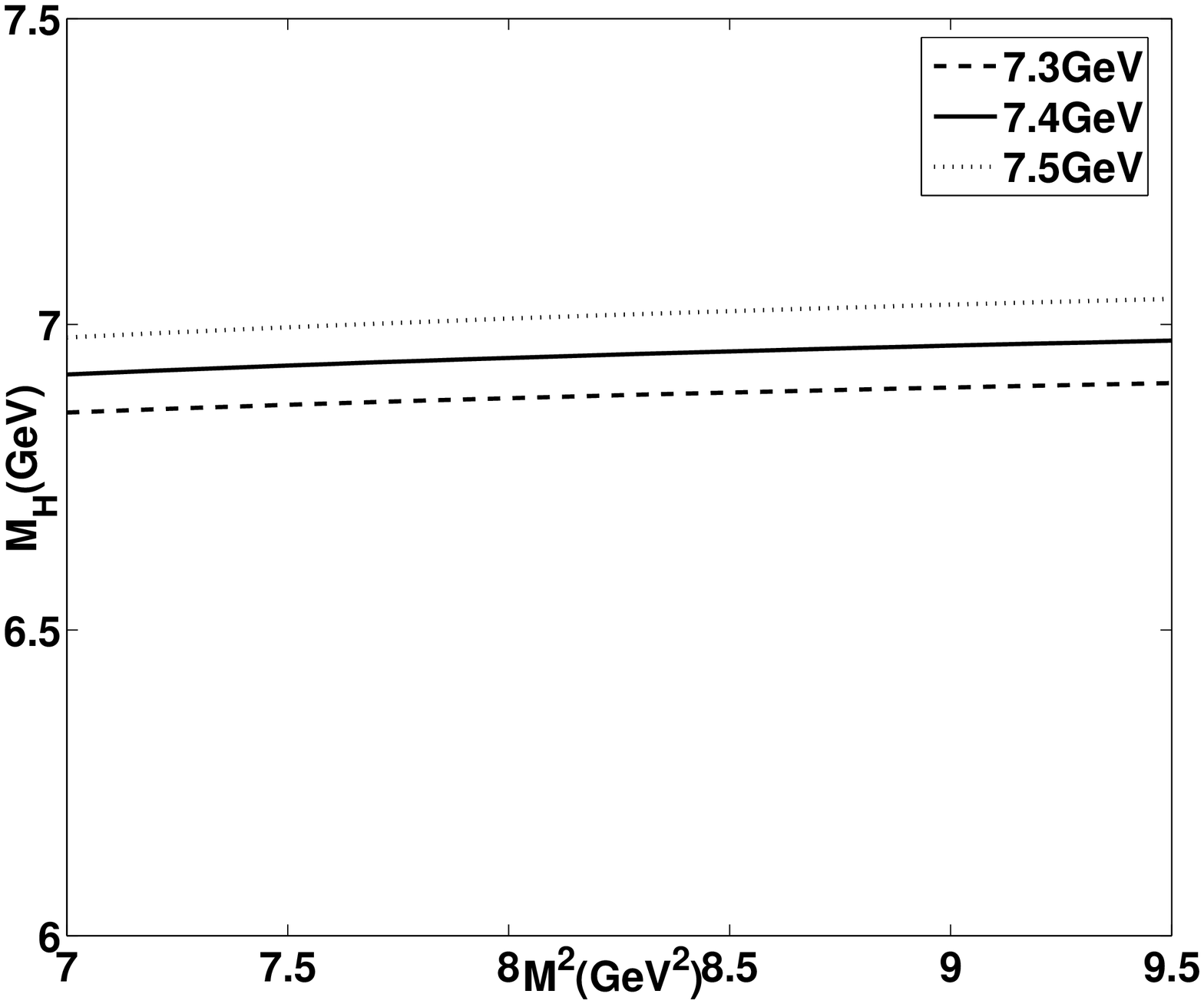}\epsfysize=3.7truecm
\epsfbox{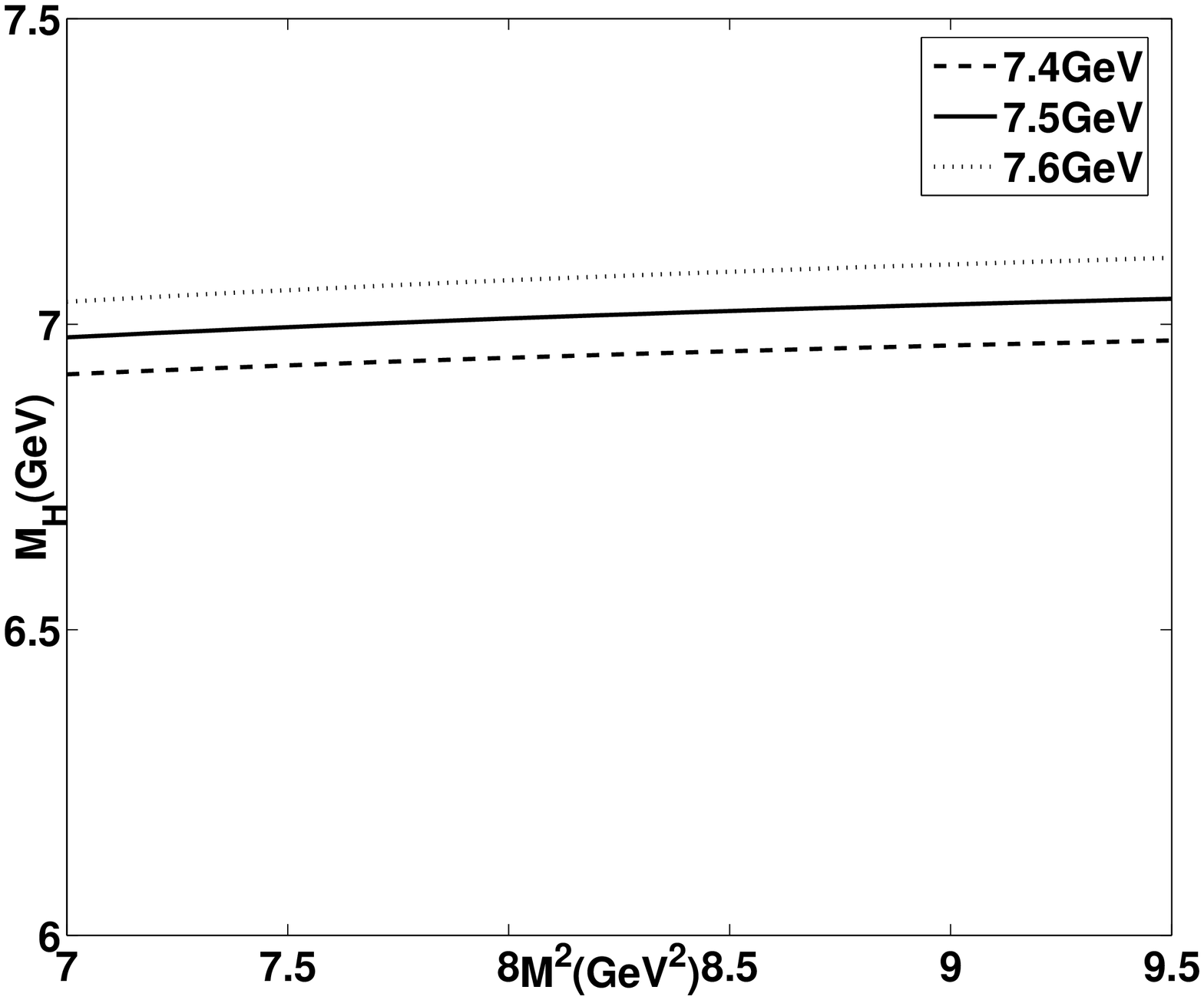}}\caption{The dependence on $M^2$ for the
masses of $\Xi_{cb}^{'}$ and $\Omega_{cb}^{'}$ from sum rule
(\ref{sum rule q}). The continuum thresholds are taken as
$\sqrt{s_0}=7.3\sim7.5~\mbox{GeV}$ and
$\sqrt{s_0}=7.4\sim7.6~\mbox{GeV}$.} \label{fig:7}
\end{figure}

\begin{table}[htb!]\caption{ The mass spectra of doubly heavy baryons (mass in
unit of$~\mbox{GeV}$).}
 \centerline{\begin{tabular}{ c  c  c  c  c  c  c  c  c c c c }  \hline\hline
Baryon                          & content                &     $J^{P}$            &  $S_{d}$     &  $L_{d}$     &   $J_{d}^{P_{d}}$         & This work             &\cite{quark model}    & \cite{MIT bag model} &\cite{mass formulas}& \cite{NRQCDSR}   &  \cite{EBagan}     \\
\hline
 $\Xi_{cc}$                     &$\{cc\}q$               &  $\frac{1}{2}^{+}$     &     1        &      0       &        $1^{+}$            & $4.26\pm0.19$         &    $3.620$           &  $3.520$             &     $3.676$        &  $3.55\pm0.08$   &   $3.48\pm0.06$    \\
\hline
 $\Xi_{cc}^{*}$                 &$\{cc\}q$               &  $\frac{3}{2}^{+}$     &     1        &      0       &        $1^{+}$            & $3.90\pm0.10$         &    $3.727$           &  $3.63$              &     $3.746$        &                  &   $3.58\pm0.05$    \\
\hline
 $\Omega_{cc}$                  &$\{cc\}s$               &  $\frac{1}{2}^{+}$     &     1        &      0       &        $1^{+}$            & $4.25\pm0.20$         &    $3.778$           &  $3.619$             &     $3.787$        &  $3.65\pm0.07$   &                    \\
\hline
 $\Omega_{cc}^{*}$              &$\{cc\}s$               &  $\frac{3}{2}^{+}$     &     1        &      0       &        $1^{+}$            & $3.81\pm0.06$         &    $3.872$           &  $3.721$             &     $3.851$        &                  &                    \\
\hline
 $\Xi_{bb}$                     &$\{bb\}q$               &  $\frac{1}{2}^{+}$     &     1        &      0       &        $1^{+}$            & $9.78\pm0.07$         &    $10.202$          &  $10.272$            &                    &  $10.00\pm0.08$  &   $9.94\pm0.91$    \\
\hline
 $\Xi_{bb}^{*}$                 &$\{bb\}q$               &  $\frac{3}{2}^{+}$     &     1        &      0       &        $1^{+}$            & $10.35\pm0.08$        &    $10.237$          &  $10.337$            &     $10.398$       &                  &   $10.33\pm1.09$   \\
\hline
 $\Omega_{bb}$                  &$\{bb\}s$               &  $\frac{1}{2}^{+}$     &     1        &      0       &        $1^{+}$            & $9.85\pm0.07$         &    $10.359$          &  $10.369$            &                    & $10.09\pm0.07$   &                    \\
\hline
 $\Omega_{bb}^{*}$              &$\{bb\}s$               &  $\frac{3}{2}^{+}$     &     1        &      0       &        $1^{+}$            & $10.28\pm0.05$        &    $10.389$          &  $10.429$            &     $10.483$       &                  &                    \\
 \hline
 $\Xi_{cb}$                     &$\{cb\}q$               &  $\frac{1}{2}^{+}$     &     1        &      0       &        $1^{+}$            & $6.75\pm0.05$         &    $6.933$           &  $6.838$             &     $7.053$        & $6.79\pm0.08$    &                    \\
\hline
 $\Xi_{cb}^{*}$                 &$\{cb\}q$               &  $\frac{3}{2}^{+}$     &     1        &      0       &        $1^{+}$            & $8.00\pm0.26$         &    $6.980$           &  $6.986$             &     $7.083$        &                  &                    \\
\hline
 $\Omega_{cb}$                  &$\{cb\}s$               &  $\frac{1}{2}^{+}$     &     1        &      0       &        $1^{+}$            & $7.02\pm0.08$         &    $7.088$           &  $6.941$             &     $7.148$        & $6.89\pm0.07$    &                    \\
\hline
 $\Omega_{cb}^{*}$              &$\{cb\}s$               &  $\frac{3}{2}^{+}$     &     1        &      0       &        $1^{+}$            & $7.54\pm0.08$         &    $7.130$           &  $7.077$             &     $7.165$        &                  &                    \\
\hline
 $\Xi_{cb}^{'}$                 &$[cb]q$                 &  $\frac{1}{2}^{+}$     &     0        &      0       &        $0^{+}$            & $6.95\pm0.08$         &    $6.963$           &  $7.028$             &     $7.062$        &                  &   $6.44\pm0.19$    \\
\hline
 $\Omega_{cb}^{'}$              &$[cb]s$                 &  $\frac{1}{2}^{+}$     &     0        &      0       &        $0^{+}$            & $7.02\pm0.08$         &    $7.116$           &  $7.116$             &     $7.151$        &                  &                    \\
\hline\hline
\end{tabular}} \label{table:2}
\end{table}
In summary, the QCD sum rules have been employed to compute the
masses of doubly heavy baryons, including the contributions of the
operators up to dimension six in OPE.  The final results are in
agreement with the existing predictions of other approaches, which
may support the heavy-diquark--light-quark configuration of doubly
heavy baryons. Anyhow, the results from all the theoretical
approaches are looking forward to experimental identification. It is
expected that the Large Hadron Collider may supply a gap of the
doubly heavy baryon data in future.

\begin{acknowledgments}
J.R.Z. is very grateful to Marina Nielsen for communications and helpful discussions.
M.Q.H. would like to thank the Abdus Salam ICTP for warm hospitality.
This work was supported in part by the National Natural Science
Foundation of China under Contract No.10675167.
\end{acknowledgments}

\end{document}